\documentclass[12pt,a4paper]{article}

\usepackage{ifthen} 
\newboolean{pdflatex}
\setboolean{pdflatex}{true} 

\newboolean{articletitles}
\setboolean{articletitles}{true} 

\newboolean{uprightparticles}
\setboolean{uprightparticles}{false} 

\newboolean{inbibliography}
\setboolean{inbibliography}{false} 
\usepackage{multirow}
\usepackage{arydshln}

\def\paperauthors{LHCb collaboration} 
\def\paperasciititle{First measurement of the ratio of psi(2S)-to-J/psi inclusive production in pAr and pp collisions at 113 GeV with SMOG2} 
\def\papertitle{First measurement of the ratio of \psitwos-to-\jpsi inclusive production in \ensuremath{p\text{Ar}} and \ensuremath{pp} collisions at $\sqsnn=113\gev$ with SMOG2}
\def\paperkeywords{{High Energy Physics}, {LHCb}} 
\def\papercopyright{2026 CERN for the benefit of the LHCb collaboration}
\def\paperlicence{CC-BY-4.0}
\def\paperlicenceurl{https://creativecommons.org/licenses/by/4.0/}

\newif\ifEnableSectionTOCLinks
\EnableSectionTOCLinksfalse 

\usepackage[top=1in, bottom=1.25in, left=1in, right=1in]{geometry}

\usepackage{microtype}
\usepackage{lineno}  
\usepackage{xspace} 
\usepackage{caption} 

\usepackage{graphicx}  
\usepackage{color}
\usepackage{colortbl}
\graphicspath{{./figs/}} 

\usepackage{amsmath} 
\usepackage{amssymb}
\usepackage{amsfonts}
\usepackage{upgreek} 

\newcommand*\patchAmsMathEnvironmentForLineno[1]{%
\expandafter\let\csname old#1\expandafter\endcsname\csname #1\endcsname
\expandafter\let\csname oldend#1\expandafter\endcsname\csname
end#1\endcsname
 \renewenvironment{#1}%
   {\linenomath\csname old#1\endcsname}%
   {\csname oldend#1\endcsname\endlinenomath}%
}
\newcommand*\patchBothAmsMathEnvironmentsForLineno[1]{%
  \patchAmsMathEnvironmentForLineno{#1}%
  \patchAmsMathEnvironmentForLineno{#1*}%
}
\AtBeginDocument{%
\patchBothAmsMathEnvironmentsForLineno{equation}%
\patchBothAmsMathEnvironmentsForLineno{align}%
\patchBothAmsMathEnvironmentsForLineno{flalign}%
\patchBothAmsMathEnvironmentsForLineno{alignat}%
\patchBothAmsMathEnvironmentsForLineno{gather}%
\patchBothAmsMathEnvironmentsForLineno{multline}%
\patchBothAmsMathEnvironmentsForLineno{eqnarray}%
}

\usepackage[pdftex,
            pdfauthor={\paperauthors},
            pdftitle={\paperasciititle},
            pdfkeywords={\paperkeywords}]{hyperref}
\usepackage{hyperxmp}
\hypersetup{
    pdfcopyright={Copyright (C) \papercopyright},
    pdflicenseurl={\paperlicenceurl}
}

\usepackage[colorinlistoftodos,textsize=scriptsize]{todonotes}

\usepackage[bottom,flushmargin,hang,multiple]{footmisc}

\usepackage[all]{hypcap} 

\usepackage{xspace} 
\usepackage{upgreek}

\def\MagUp {\mbox{\em Mag\kern -0.05em Up}\xspace}

\ifdefined\ifuprightparticles
\else
\newboolean{uprightparticles}
\setboolean{uprightparticles}{false} 
\fi

\ifthenelse{\boolean{uprightparticles}}%
{

 \def\Pmu         {\ensuremath{\upmu}\xspace}

 \def\Ppsi        {\ensuremath{\uppsi}\xspace}

 \def\PDelta      {\ensuremath{\Delta}\xspace}                 
 \def\PXi         {\ensuremath{\Xi}\xspace}                 
 \def\PLambda     {\ensuremath{\Lambda}\xspace}                 
 \def\PSigma      {\ensuremath{\Sigma}\xspace}                 
 \def\POmega      {\ensuremath{\Omega}\xspace}                 
 \def\PUpsilon    {\ensuremath{\Upsilon}\xspace}
 \let\oldPi\Pi
 \def\PPi         {\ensuremath{\oldPi}\xspace}

 \def\PB      {\ensuremath{\mathrm{B}}\xspace}                 
 \def\PD      {\ensuremath{\mathrm{D}}\xspace}                 
 \def\PJ      {\ensuremath{\mathrm{J}}\xspace}                 
 \def\PK      {\ensuremath{\mathrm{K}}\xspace}                 
 \def\Pb      {\ensuremath{\mathrm{b}}\xspace}                 
 \def\Pc      {\ensuremath{\mathrm{c}}\xspace}

 \def\Ps      {\ensuremath{\mathrm{s}}\xspace}

 \def\thebaroffset{0.0em}
}
{

 \def\Pmu         {\ensuremath{\mu}\xspace}

 \def\Ppsi        {\ensuremath{\psi}\xspace}                 
                  
 \mathchardef\PDelta="7101
 \mathchardef\PXi="7104
 \mathchardef\PLambda="7103
 \mathchardef\PSigma="7106
 \mathchardef\POmega="710A
 \mathchardef\PUpsilon="7107
 \mathchardef\PPi="7105
 \def\PB      {\ensuremath{B}\xspace}                 
 \def\PD      {\ensuremath{D}\xspace}                 
 \def\PJ      {\ensuremath{J}\xspace}                 
 \def\PK      {\ensuremath{K}\xspace}                 
 \def\Pb      {\ensuremath{b}\xspace}                 
 \def\Pc      {\ensuremath{c}\xspace}

 \def\Ps      {\ensuremath{s}\xspace}

 \def\thebaroffset{0.18em}
}
\newcommand{\offsetoverline}[2][\thebaroffset]{\kern #1\overline{\kern -#1 #2}}%

\makeatletter
\ifcase \@ptsize \relax
  \newcommand{\miniscule}{\@setfontsize\miniscule{4}{5}}
\or
  \newcommand{\miniscule}{\@setfontsize\miniscule{5}{6}}
\or
  \newcommand{\miniscule}{\@setfontsize\miniscule{5}{6}}
\fi
\makeatother

\DeclareRobustCommand{\optbar}[1]{\shortstack{{\miniscule (\rule[.5ex]{1.25em}{.18mm})}
  \\ [-.7ex] $#1$}}

\def\mup        {{\ensuremath{\Pmu^+}}\xspace}
\def\mun        {{\ensuremath{\Pmu^-}}\xspace} 

\def\squark    {{\ensuremath{\Ps}}\xspace}

\def\cquark    {{\ensuremath{\Pc}}\xspace}
\def\cquarkbar {{\ensuremath{\overline \cquark}}\xspace}
\def\ccbar     {{\ensuremath{\cquark\cquarkbar}}\xspace}
\def\bquark    {{\ensuremath{\Pb}}\xspace}
\def\bquarkbar {{\ensuremath{\overline \bquark}}\xspace}
\def\bbbar     {{\ensuremath{\bquark\bquarkbar}}\xspace}

\def\kaon    {{\ensuremath{\PK}}\xspace}

\def\KorKbar {\kern \thebaroffset\optbar{\kern -\thebaroffset \PK}{}\xspace}

\def\Kp      {{\ensuremath{\kaon^+}}\xspace}
\def\Km      {{\ensuremath{\kaon^-}}\xspace}

\def\D       {{\ensuremath{\PD}}\xspace}

\def\DorDbar {\kern \thebaroffset\optbar{\kern -\thebaroffset \PD}\xspace}

\def\Dp      {{\ensuremath{\D^+}}\xspace}
\def\Dm      {{\ensuremath{\D^-}}\xspace}

\def\DpDm    {\ensuremath{\Dp {\kern -0.16em \Dm}}\xspace}

\def\B       {{\ensuremath{\PB}}\xspace}

\def\BorBbar {\kern \thebaroffset\optbar{\kern -\thebaroffset \PB}\xspace}

\def\Bd      {{\ensuremath{\B^0}}\xspace}

\def\BdorBdbar {\kern \thebaroffset\optbar{\kern -\thebaroffset \Bd}\xspace}

\def\Bs      {{\ensuremath{\B^0_\squark}}\xspace}

\def\BsorBsbar {\kern \thebaroffset\optbar{\kern -\thebaroffset \Bs}\xspace}

\def\jpsi     {{\ensuremath{{\PJ\mskip -3mu/\mskip -2mu\Ppsi}}}\xspace}
\def\psitwos  {{\ensuremath{\Ppsi{(2S)}}}\xspace}

\def\Y#1S{\ensuremath{\PUpsilon{(#1S)}}\xspace}

\def\LorLbar     {\kern \thebaroffset\optbar{\kern -\thebaroffset \PLambda}\xspace}

\def\BF         {{\ensuremath{\mathcal{B}}}\xspace}

\def\to                 {\ensuremath{\rightarrow}\xspace}

\def\AT#1     {\ensuremath{A_{\mathrm{T}}^{#1}}\xspace}           

\def\C#1      {\ensuremath{\mathcal{C}_{#1}}\xspace}                       
\def\Cp#1     {\ensuremath{\mathcal{C}_{#1}^{'}}\xspace}                    
\def\Ceff#1   {\ensuremath{\mathcal{C}_{#1}^{\mathrm{(eff)}}}\xspace}        
\def\Cpeff#1  {\ensuremath{\mathcal{C}_{#1}^{'\mathrm{(eff)}}}\xspace}       
\def\Ope#1    {\ensuremath{\mathcal{O}_{#1}}\xspace}                       
\def\Opep#1   {\ensuremath{\mathcal{O}_{#1}^{'}}\xspace}                    

\newcommand{\aunit}[1]{\ensuremath{\text{\,#1}}}       

\newcommand{\tev}{\aunit{Te\kern -0.1em V}\xspace}
\newcommand{\gev}{\aunit{Ge\kern -0.1em V}\xspace}
\newcommand{\mev}{\aunit{Me\kern -0.1em V}\xspace}
\newcommand{\kev}{\aunit{ke\kern -0.1em V}\xspace}
\newcommand{\ev}{\aunit{e\kern -0.1em V}\xspace}
 
\newcommand{\mevc}{\ensuremath{\aunit{Me\kern -0.1em V\!/}c}\xspace}
\newcommand{\gevc}{\ensuremath{\aunit{Ge\kern -0.1em V\!/}c}\xspace}
\newcommand{\mevcc}{\ensuremath{\aunit{Me\kern -0.1em V\!/}c^2}\xspace}
\newcommand{\gevcc}{\ensuremath{\aunit{Ge\kern -0.1em V\!/}c^2}\xspace}

\def\mm   {\aunit{mm}\xspace}

\def\mbarn{\aunit{mb}\xspace}

\def\pb {\aunit{pb}\xspace}
\def\invpb {\ensuremath{\pb^{-1}}\xspace}

\def\gsim{{~\raise.15em\hbox{$>$}\kern-.85em
          \lower.35em\hbox{$\sim$}~}\xspace}
\def\lsim{{~\raise.15em\hbox{$<$}\kern-.85em
          \lower.35em\hbox{$\sim$}~}\xspace}

\def\sPlot{\mbox{\em sPlot}\xspace}

\def\sqs   {\ensuremath{\protect\sqrt{s}}\xspace}
\def\sqsnn {\ensuremath{\protect\sqrt{s_{\scriptscriptstyle\text{NN}}}}\xspace}
\def\pt         {\ensuremath{p_{\mathrm{T}}}\xspace}

\def\ptot       {\ensuremath{p}\xspace}

\def\evtgen     {\mbox{\textsc{EvtGen}}\xspace}

\def\geant      {\mbox{\textsc{Geant4}}\xspace}

\def\photos     {\mbox{\textsc{Photos}}\xspace}

\def\pythia     {\mbox{\textsc{Pythia}}\xspace}

\def\tell1  {TELL1\xspace}
\def\ukl1   {UKL1\xspace}

\newcommand{\phz}{\phantom{0}}
\newcommand{\phm}{\phantom{-}}
\newcommand{\lhcborcid}[1]{\href{https://orcid.org/#1}{\hspace*{0.1em}\raisebox{-0.45ex}{\includegraphics[width=1em]{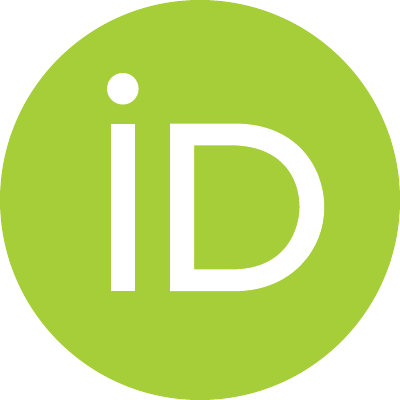}}}}

\hypersetup{
  colorlinks   = true, 
  urlcolor     = blue, 
  linkcolor    = blue, 
  citecolor    = red   
}

\ifEnableSectionTOCLinks
    \usepackage[explicit]{titlesec} 
    
    \let\oldcontentsline\contentsline
    \renewcommand

    \titleformat{\section}{\normalfont\Large\bf}{\hyperlink{tocsection.\thesection}{{\thesection} \parbox[t]{\dimexpr\textwidth-1pc}{#1}}}{1pc}{}

    \titleformat{\subsection}{\normalfont\bf}{\hyperlink{tocsubsection.\thesubsection}{{\thesubsection} \parbox[t]{\dimexpr\textwidth-1pc}{#1}}}{1pc}{}

    \titleformat{name=\section,numberless}[display]{}{}{0pt}{\normalfont\Huge\bfseries #1}
\fi

\usepackage{cite} 
\usepackage{mciteplus}

\newcommand{\pp}{\ensuremath{pp}\xspace}

\newcommand{\pAr}{\ensuremath{p\text{Ar}}\xspace}

\newcommand*{\pAl}{\ensuremath{p\mathrm{Al}}\,}
\newcommand*{\pAu}{\ensuremath{p\mathrm{Au}}\,}

\newcommand*{\HethreeAu}{\ensuremath{{}^{3}\mathrm{He}\mathrm{Au}}\,}

\newcommand{\Tabref}[1]{Table~\protect\ref{#1}}

\newcommand{\Figref}[1]{Fig.~\protect\ref{#1}}

\newcommand{\Appref}[1]{Appendix~\protect\ref{#1}}

\newcommand{\Eqref}[1]{Eq.~\protect\ref{#1}}

\def\ystar {\ensuremath{y^{*}\,\xspace}}

\begin{document}

\renewcommand{\thefootnote}{\fnsymbol{footnote}}
\setcounter{footnote}{1}



\begin{titlepage}
\pagenumbering{roman}
\vspace*{-1.5cm}
\centerline{\large EUROPEAN ORGANIZATION FOR NUCLEAR RESEARCH (CERN)}
\vspace*{1.5cm}
\noindent
\begin{tabular*}{\linewidth}{lc@{\extracolsep{\fill}}r@{\extracolsep{0pt}}}
\ifthenelse{\boolean{pdflatex}}
{\vspace*{-1.5cm}\mbox{\!\!\!\includegraphics[width=.14\textwidth]{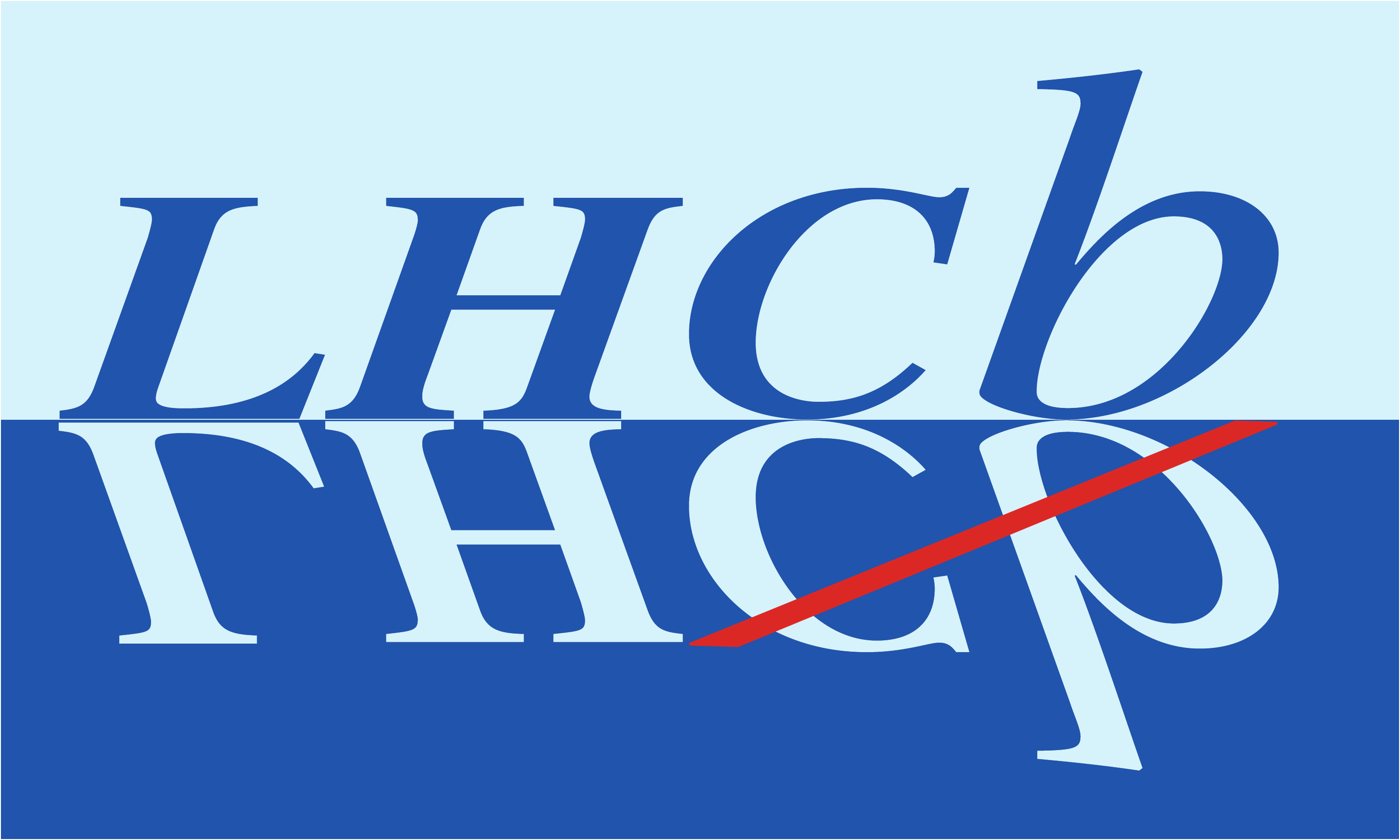}} & &}%
{\vspace*{-1.2cm}\mbox{\!\!\!\includegraphics[width=.12\textwidth]{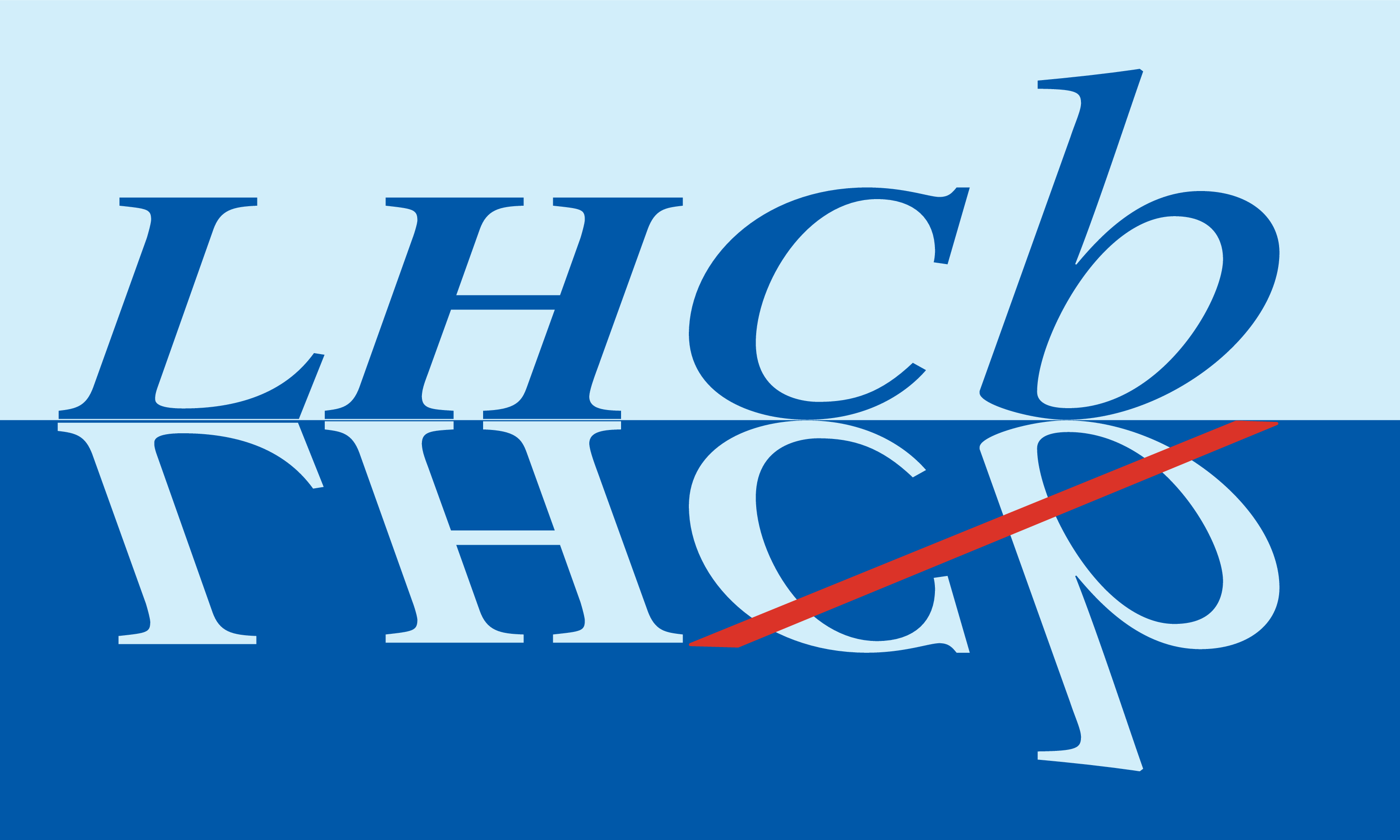}} & &}%
\\
 & & CERN-EP-2026-242 \\  
 & & LHCb-PAPER-2026-026  \\  
 & & 31 August 2026 \\ 
 & & \\
\end{tabular*}
\vspace*{2.0cm}
{\normalfont\bfseries\boldmath\huge
\begin{center}
  \papertitle 
\end{center}
}
\begin{center}
\paperauthors\footnote{Authors are listed at the end of this paper.}
\end{center}
\vspace{\fill}
\vspace*{1.0cm}
\begin{abstract}
  \noindent
  A measurement of the $\psitwos$-to-$\jpsi$ production cross-section ratio is performed in proton-argon $ (\pAr$) and proton-proton ($pp$) collisions in fixed-target mode at \mbox{$\sqsnn=113\gev$}. Data samples were collected by the LHCb experiment during argon and hydrogen gas injections in the SMOG2 storage cell, resulting in $\pAr$ and $\pp$ collisions, respectively. The $\psitwos$-to-$\jpsi$ production cross-section ratio is measured as a function of the charmonium transverse momentum, $\pt$, and rapidity in the centre-of-mass system, $y^{*}$. The $\psitwos$-to-$\jpsi$ ratio in $\pAr$ collisions over that in \pp collisions is measured to be $0.90 \pm 0.04 \pm 0.02$ for $-2.3<\ystar<0.0$ and $0<\pt<8\gevc$, indicating the emergence of nuclear effects in the $\pAr$ system. This study acts as a baseline for the interpretation of future measurements with larger systems accessible by the LHCb experiment.
\end{abstract}
\vspace*{1.0cm}
\begin{center}
Submitted to
  JHEP 
\end{center}
\vspace{\fill}
{\footnotesize 
\centerline{\copyright~\papercopyright. \href{\paperlicenceurl}{\paperlicence} licence.}}
\vspace*{2mm}
\end{titlepage}


\newpage
\setcounter{page}{2}
\mbox{~}
\cleardoublepage
\renewcommand{\thefootnote}{\arabic{footnote}}
\setcounter{footnote}{0}


\pagestyle{plain} 
\setcounter{page}{1}
\pagenumbering{arabic}


\section{Introduction}
\label{sec:Introduction}

The production of charmonia, bound states of a charm quark and its antiquark (\ccbar), in proton–proton (\pp), proton–nucleus, and nucleus–nucleus collisions is an important process to study Quantum Chromodynamics (QCD). This process involves at least two distinct scales: the hard scale to produce the heavy \ccbar pair and the binding energy of the bound state. In proton–nucleus collisions, several initial- and final-state effects can alter charmonium production compared to \pp collisions. Charmonium production may be suppressed by nuclear absorption \cite{NAPLB207}, and affected by multiple scattering and radiative energy loss in the proton–nucleus overlapping region \cite{ELJHEP03}. Furthermore, charmonium states can be dissociated by comoving particles (comovers) \cite{CapellaPRL85} or influenced by modifications of the parton flux within the nucleus \cite{Eskola:2021nhw,AbdulKhalek:2022fyi}. These effects depend on several factors: the collision energy, the transverse momentum and rapidity of the charmonium state, and the size of the target nucleus. Apart from proton-nucleus collisions, these phenomena are also observed in larger systems such as heavy nucleus-nucleus collisions, where additional effects associated with the formation of a Quark Gluon Plasma are expected~\cite{Rothkopf:2019ipj}. This complex picture can only be fully understood with precision charmonium measurements across a broad range of experimental conditions.

A key observable to probe final-state effects is the relative production rate of excited and low-lying charmonium states, such as the \psitwos-to-\jpsi ratio. Since the \psitwos resonance is more weakly bound than the \jpsi meson, it is expected to be more sensitive to suppression from final-state effects~\cite{CapellaPRL85}.
In this analysis, a measurement of inclusive charmonium production by the LHCb experiment in its fixed-target configuration is presented. The production cross-section ratio of \psitwos and \jpsi mesons is studied in collisions of $6800\gev$ protons with hydrogen and argon gas targets at rest, resulting in proton-argon (\pAr) and $pp$ collisions at a nucleon-nucleon centre-of-mass energy of $\sqsnn=113\gev$.

The \jpsi and \psitwos states are reconstructed through their decays into two oppositely charged muons. The cross-section ratio multiplied by the ratio of branching fractions is measured as
\begin{equation}
R_{\psitwos/\jpsi} \equiv \frac{\sigma_{\psitwos}}{\sigma_{\jpsi}} \frac{\BF_{\psitwos\rightarrow \mup\mun}}{\BF_{\jpsi\rightarrow \mup\mun}} = \frac{N_{\psitwos}}{N_{\jpsi}}\frac{\varepsilon_{\jpsi}}{\varepsilon_{\psitwos}},
\label{eq:xsecratiodeff}
\end{equation}
where $N_{\psitwos,\jpsi}$ are the yields, $\varepsilon_{\psitwos,\jpsi}$ are the total efficiencies and \mbox{$\BF$} are the branching fractions of \psitwos and \jpsi mesons decaying into a $\mup\mun$ pair. The use of the same final state of both states leads to the cancellation of many experimental effects in the ratio measurement. Direct information regarding nuclear effects present in \pAr collisions can be accessed through the double ratio, defined as
\begin{equation}
R^{\pAr/\pp}_{\psitwos/\jpsi} \equiv \dfrac{\left[\sigma_{\psitwos}/\sigma_{\jpsi}\right]_{\pAr}}{\left[\sigma_{\psitwos}/\sigma_{\jpsi}\right]_{\pp}}.
\label{eq:dxsecratio} 
\end{equation}

\section{Detector, datasets, event selection and simulation }
\label{sec:Detector}

The LHCb Run 3 detector~\cite{LHCb-DP-2022-002} is a single-arm forward spectrometer covering the pseudorapidity range $2 < \eta < 5$, designed for the study of particles containing $c$ or $b$ quarks. The detector elements that are particularly relevant to this analysis include a silicon-pixel vertex locator (VELO) surrounding the beam-beam interaction region that allows \cquark and \bquark hadrons to be identified from their characteristic flight distance, a high-precision tracking system that provides a measurement of the momentum, \ptot, of charged particles, and a muon detector composed of alternating layers of iron and multiwire proportional chambers~\cite{LHCb-DP-2025-007}.

LHCb has the unique capability among LHC experiments to be operated as a fixed-target experiment using the SMOG2 system~\cite{LHCb-DP-2024-002}, by recording collisions between beam particles and a gaseous target. Most of the beam-gas collisions occur within the SMOG2 gas storage cell, located approximately $340\mm$ upstream of the nominal beam-beam interaction region. During LHC Run 3, fixed-target collisions with various gas-target species are recorded simultaneously with beam-beam collisions.

The data samples used in this analysis were taken in 2024 while injecting argon and hydrogen gas in SMOG2, with protons circulating the LHC at an energy of $6800\gev$, resulting in beam-gas \pAr and \pp collisions at a centre-of-mass energy per nucleon of $\sqsnn=113\gev$. Argon and hydrogen gases consist of isotopic mixtures corresponding their natural abundances. These data were taken simultaneously with \pp collisions at $\sqs=13.6\tev$. Only events in which two non-empty proton bunches were crossing at the LHCb beam-beam interaction region are used in this measurement. Data are divided into eight blocks according to data-taking periods with substantially different trigger and detector spatial alignment conditions, as defined in Ref.~\cite{LHCb-PAPER-2025-036}. Data used in this analysis correspond to blocks D and E. In block D, samples of \pp and \pAr collisions with integrated luminosities of $2.4\invpb$ and $0.34\invpb$, respectively, are considered. Another larger \pp data sample from block E, corresponding to an integrated luminosity of $6.2\invpb$, is also analysed. A summary of the data samples used in this measurement and corresponding integrated luminosities is presented in \Tabref{tab:samples}.

\begin{table}[t]
\caption{ Data samples used in this analysis and corresponding integrated luminosities. }
\label{tab:samples}
\centering
\begin{tabular}{cccc}
Data block & Injected gas & Collision          & Integrated luminosity $[\!\invpb]$ \\ \hline
D          & Argon        & \pAr &                      0.34          \\
D          & Hydrogen     & \pp  &                       2.4          \\
E          & Hydrogen     & \pp  &                       6.2           
\end{tabular}
\end{table}

Online event selections are performed by a two-stage all-software trigger system~\cite{LHCb-TDR-016,LHCb-TDR-021}: a GPU-based inclusive stage named High Level Trigger 1 (HLT1) followed by a CPU-based stage (HLT2). The HLT1 stage is focused primarily on charged particle reconstruction, which reduces the data volume by roughly a factor of thirty, while HLT2  performs the full offline-quality reconstruction and selection of physics signatures. Events of beam-gas collisions are selected with dedicated trigger algorithms in HLT1 and HLT2.

Events are selected in HLT1 by requiring two oppositely charged muons with \mbox{$\ptot>3\gevc$} and transverse momentum $\pt>0.5\gevc$. The two muons are required to form a charmonium candidate with a decay vertex in the SMOG2 storage cell region and an invariant mass above $2.7\gevcc$. In HLT2, charmonium candidates associated to the primary vertex (PV) in the SMOG2 region and with an invariant mass within $[2.9,4.0]\gevcc$ are retained for further analysis. Triggered data further undergo centralised, offline processing steps to deliver physics-analysis-ready data across the entire LHCb physics programme~\cite{Sprucing,FunTuple}. For the offline selection, muons are required to be positively identified and have $\ptot>5\gevc$ and $\pt>0.6\gevc$.

The measurement is performed as a function of the charmonium \pt and rapidity in the nucleon-nucleon centre-of-mass system, \ystar, in the range $0<\pt<8\gevc$ and $-2.3<\ystar<0.0$. Due to the boost induced by the high-energy proton beam, the LHCb acceptance covers the backward rapidity hemisphere in the nucleon-nucleon centre-of-mass system, where positive rapidity is defined along the incident proton beam direction. The measurement is reported for different single and double-differential $(\pt,\ystar\!)$ intervals to investigate the kinematic dependence of potentially emerging nuclear effects. The intervals are listed in \Tabref{tab:kinematicbins}.

\begin{table}[t]
\caption{ Intervals of \pt and \ystar  in which the measurement is performed. Additionally, the measurement is reported integrated in the $0<\pt <8\gevc$ and $-2.3<\ystar<0.0$ range. }
\label{tab:kinematicbins}
\centering
\begin{tabular}{ccc}
\pt [\gevc]   & \ystar & (\ystar,\;\pt  [\gevc]) \\  \hline
$0.0<\pt<1.5$   & $-2.3<\ystar<-1.3$            &  ($-2.3<\ystar<-0.8,\;0.0<\pt<1.5$)  \\
$1.5<\pt<3.0$   & $-1.3<\ystar<-0.6$            &  ($-2.3<\ystar<-0.8,\;1.5<\pt<8.0$)  \\
$3.0<\pt<8.0$   & $-0.6<\ystar<\phantom{-}0.0$  &  ($-0.8<\ystar<\phantom{-}0.0,\;0.0<\pt<1.5$)     \\
                &                               &  ($-0.8<\ystar<\phantom{-}0.0,\;1.5<\pt<8.0)$     \\
\end{tabular}
\end{table}

Simulation samples are used to estimate the acceptance and detection efficiency ratios and to model the charmonia mass distributions. In the simulation, prompt \jpsi and \psitwos mesons, originating from the primary beam-gas interaction, are generated using \pythia~8~\cite{Sjostrand:2007gs} with a specific LHCb configuration~\cite{LHCb-PROC-2010-056}. The generated \jpsi and \psitwos meson decay products are embedded into minimum-bias \pAr and \pp events, which are generated with the EPOS event generator~\cite{EPOS_LHC}. Decays of unstable particles are described by \evtgen~\cite{Lange:2001uf}, in which final-state radiation is generated using \photos~\cite{davidson2015photos}. Minimum-bias \pp collisions at $\sqs=13.6\tev$, occurring in the beam-beam interaction region, are simulated with \pythia 8 to reproduce the experimental detector occupancy. The interaction of the generated particles with the detector, and its response, are implemented using the \geant toolkit~\cite{Allison:2006ve}, as described in Ref.~\cite{Agostinelli:2002hh}.

There are no measurements of \jpsi and \psitwos polarisation in \pAr and \pp collisions at $\sqsnn=113\gev$. The charmonium polarisation measurement by LHCb in \pp collisions at $\sqs=7\tev$ is consistent with zero~\cite{LHCb-PAPER-2013-008,LHCb-PAPER-2013-067}. The PHENIX experiment has also obtained zero $\jpsi$ polarisation in $\pp$ collisions at $\sqs=200\gev$~\cite{PHENIX:2020dqu}. Therefore, zero \psitwos and \jpsi polarisation is assumed in this measurement and no associated systematic uncertainty is considered, following previous analyses~\cite{LHCb-PAPER-2023-024,LHCb-PAPER-2023-035,PHENIX:2016vmz}.

\section{Signal yield and efficiency determination}
\label{sec:SignalExtractionEfficiency}

\begin{figure}[t]
\centering
\includegraphics[width=0.49\linewidth]{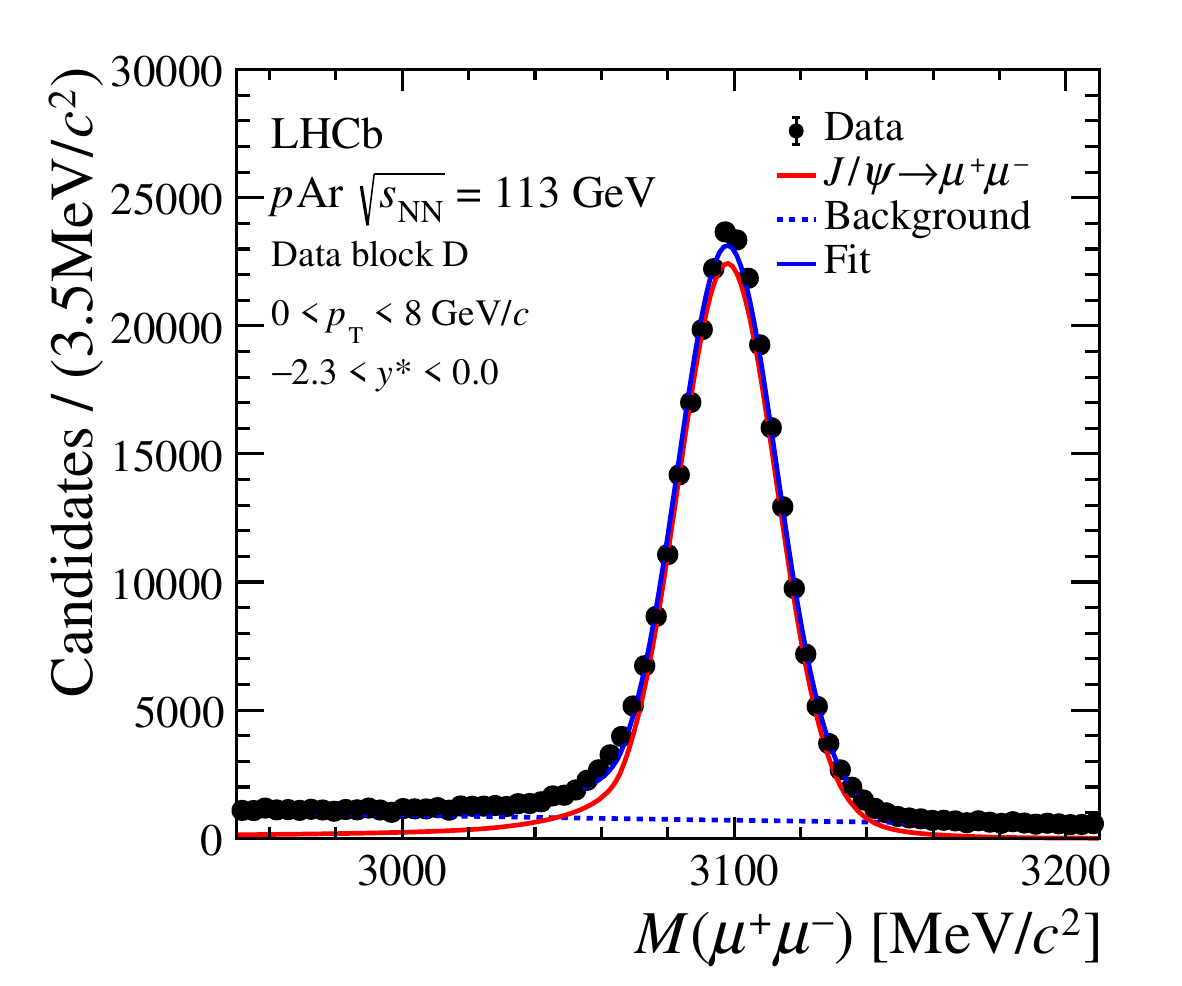}
\includegraphics[width=0.49\linewidth]{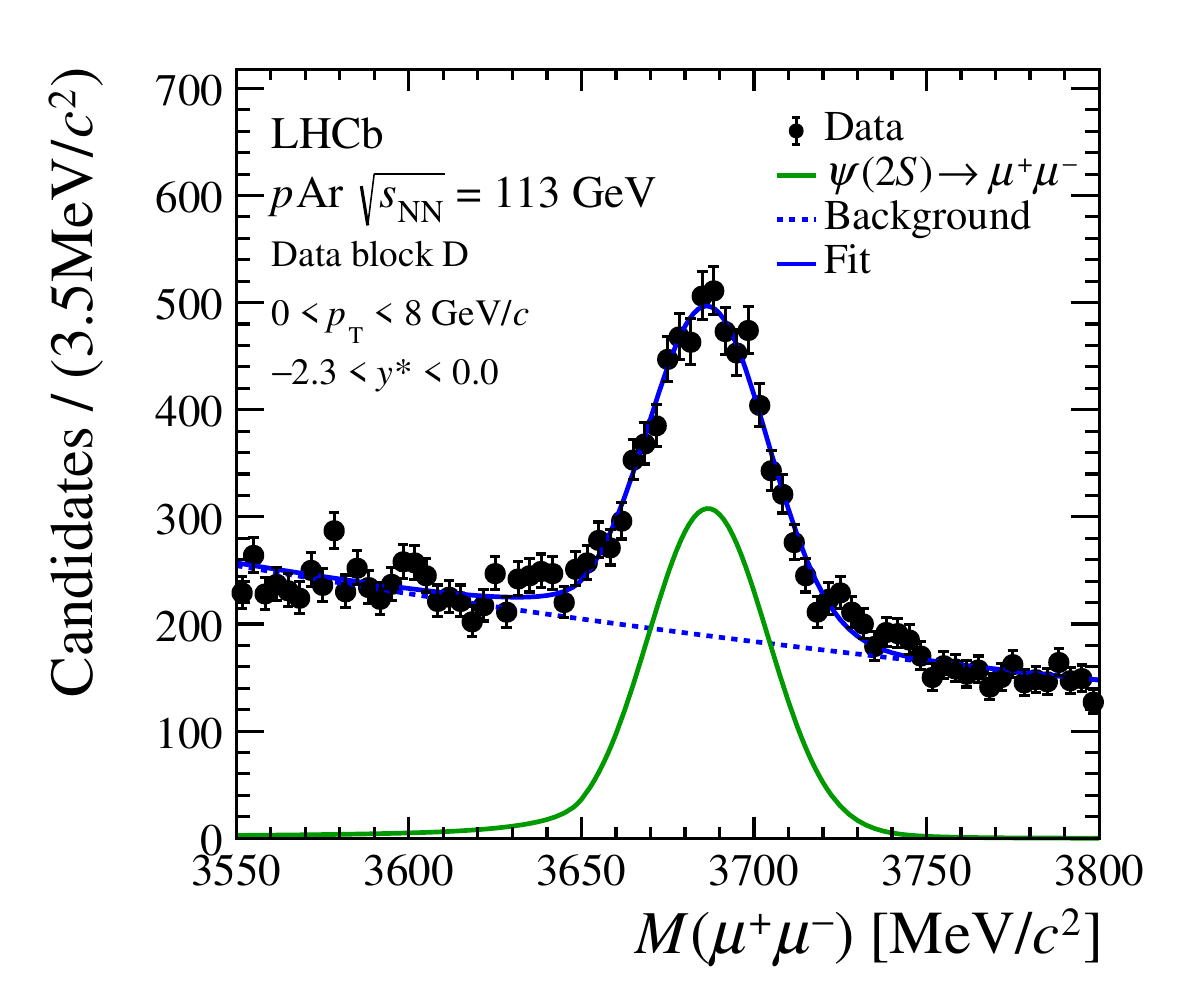}

\includegraphics[width=0.49\linewidth]{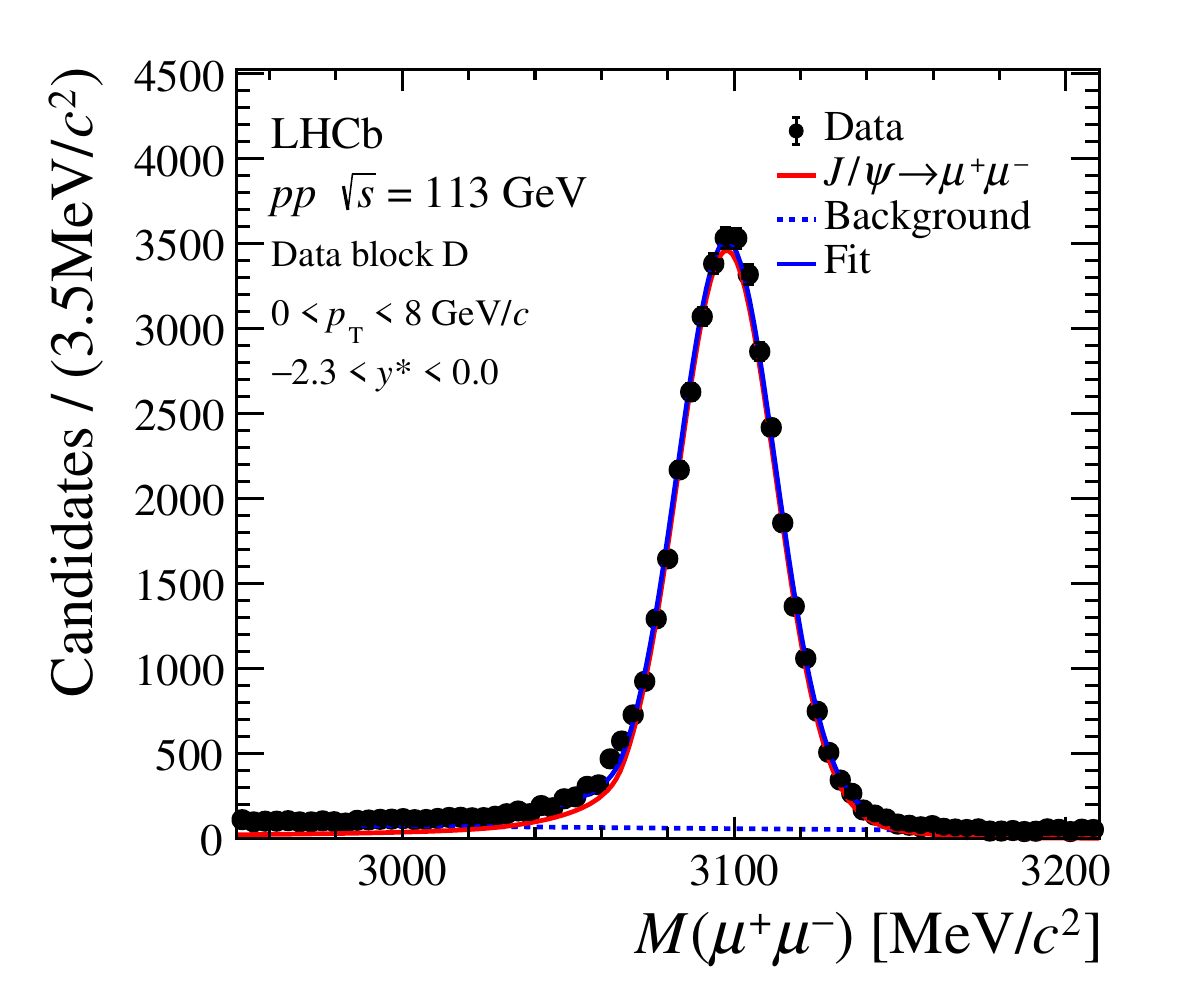}
\includegraphics[width=0.49\linewidth]{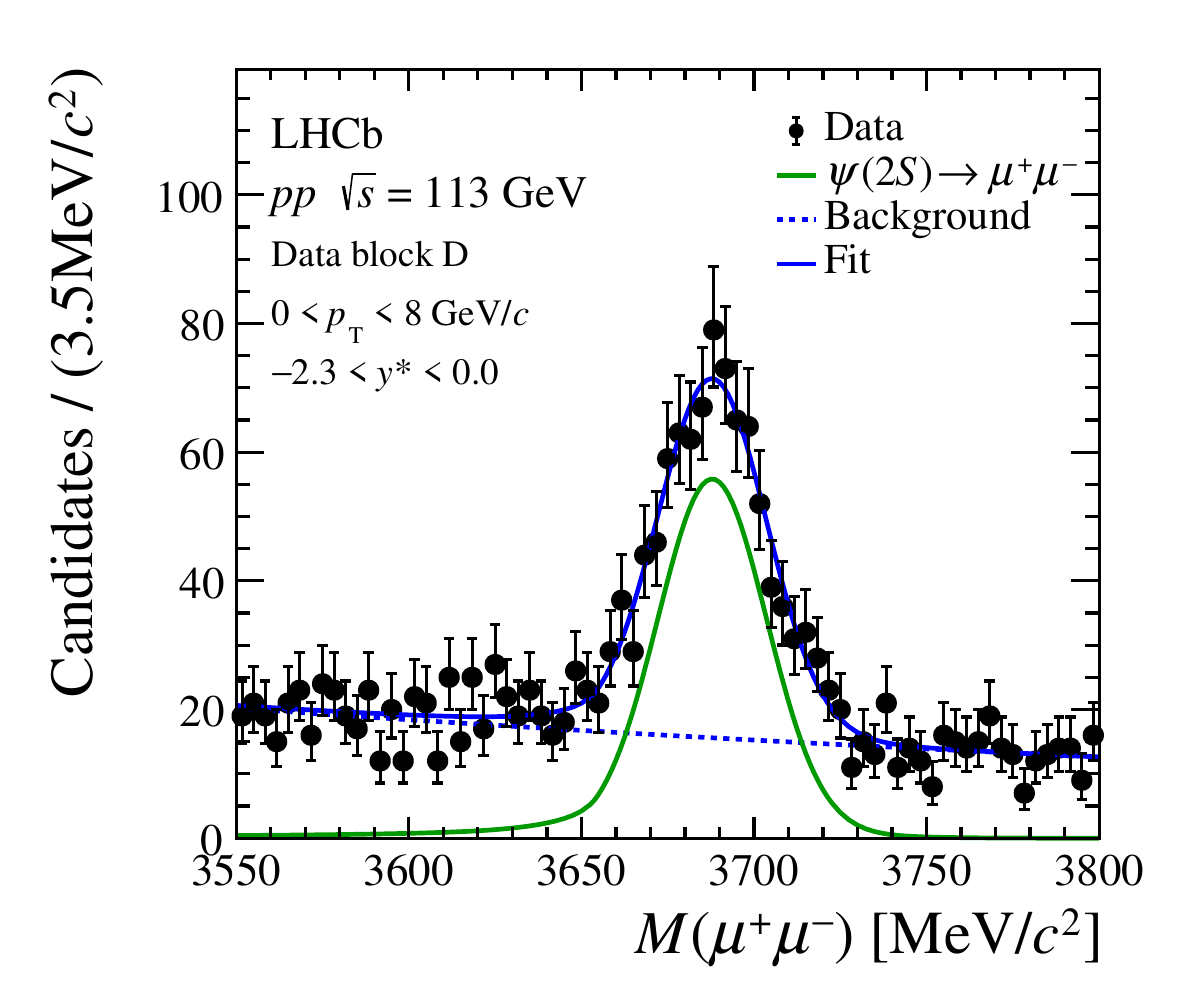}

\includegraphics[width=0.49\linewidth]{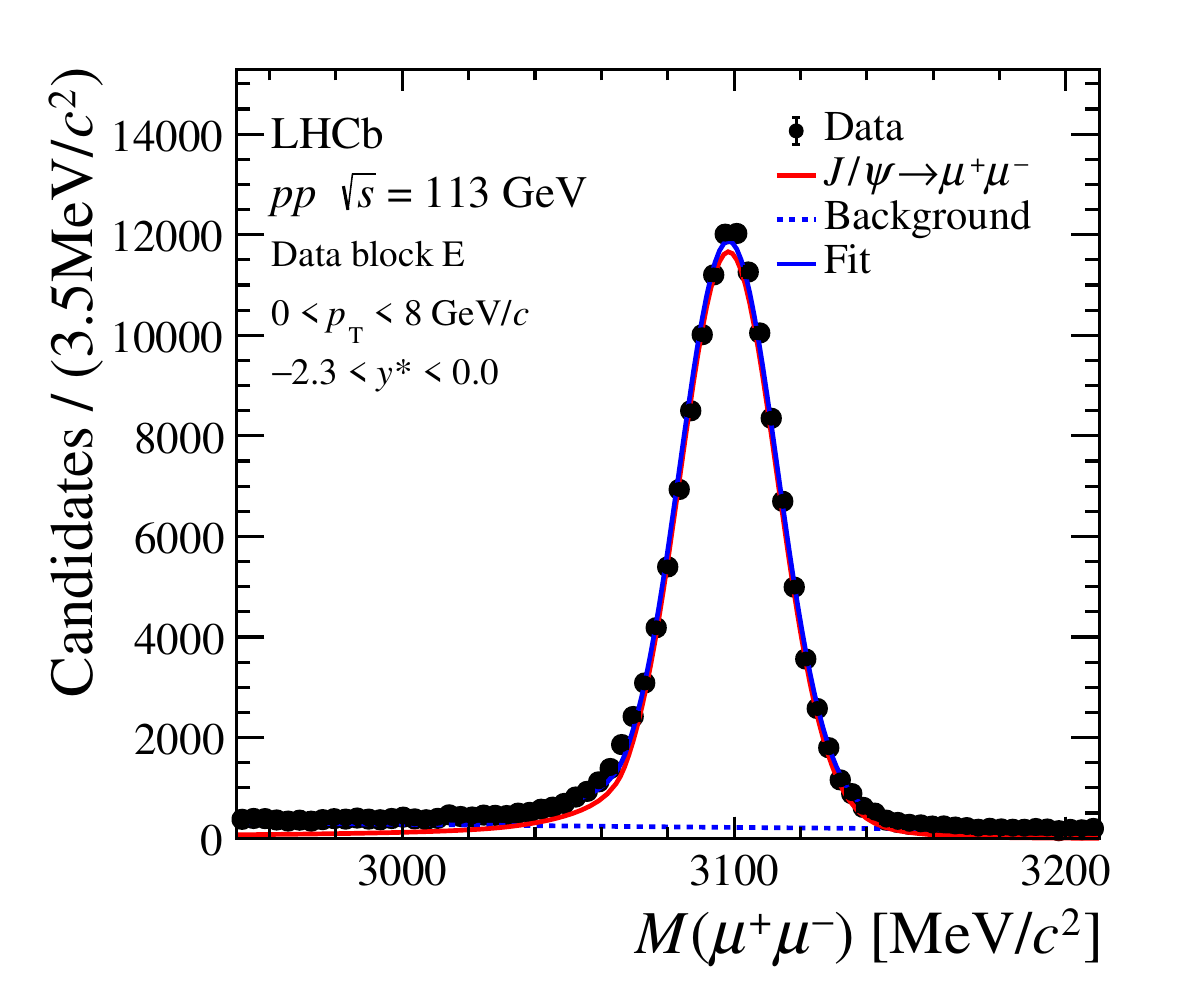}
\includegraphics[width=0.49\linewidth]{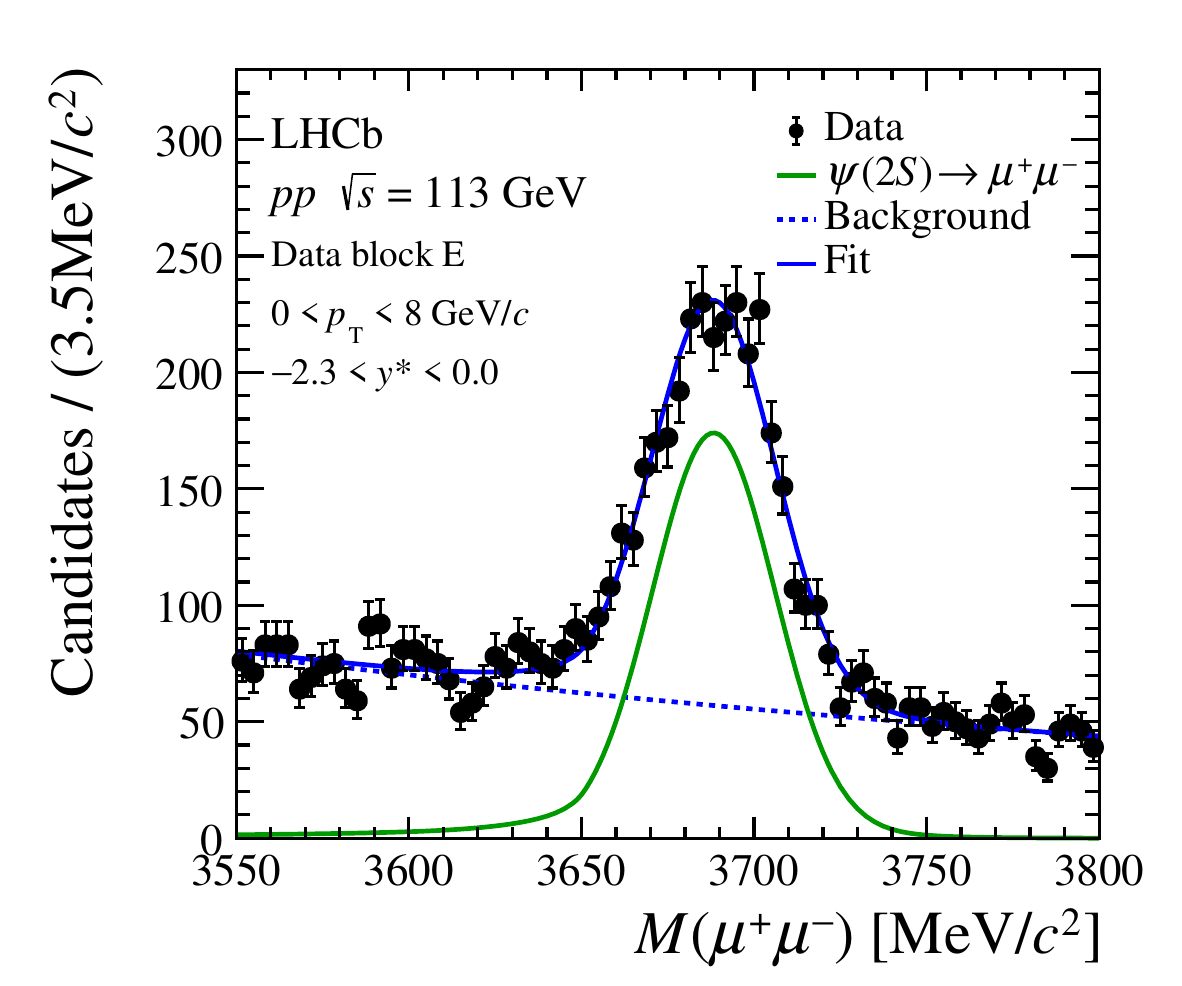}

\caption{ Mass distributions of $\mup\mun$ candidates in the (left) \jpsi and (right) \psitwos regions, together with the fit results. Top figures correspond to \pAr  data in Block D, middle figures to \pp  data in Block D and bottom figures to \pp  data in Block E.
}
\label{fig:JpsiPsi2sSignals}
\end{figure}

Signal yields of \jpsi and \psitwos mesons are obtained with extended maximum-likelihood fits to the $\mup\mun$ invariant mass in the $[2.95,3.21]\gevcc$ and $[3.55,3.80]\gevcc$ intervals for the \jpsi and \psitwos mesons, respectively. The charmonium signals are described by the sum of two Crystal Ball (CB) functions~\cite{Skwarnicki:1986xj}. The tail parameters of each CB, the relative normalisations and the relation between their widths are constrained with parametrisations extracted from fits to the simulated samples. The width of the narrower CB and the mean, which is shared by both CB functions, are free in the fit. The background is modelled by an exponential function. The fit is performed separately in each kinematic range. Figure~\ref{fig:JpsiPsi2sSignals} shows the invariant-mass distributions, together with the fit results for \jpsi and \psitwos mesons in the full kinematic range. The yields of \jpsi and \psitwos mesons, and their ratio, in each data sample, are presented in \Tabref{tab:yieldsresults}.

\begin{table}[t!]
\caption{ Yields of \jpsi, \psitwos mesons and their ratio for each sample of the analysis. Uncertainties are statistical only.}
\label{tab:yieldsresults}
\centering
\begin{tabular}{ccccc}
Data Block &  Collisions  &  $N_{\jpsi}$                &  $N_{\psitwos}$                       &  $[N_{\psitwos}/N_{\jpsi}]\times 10^{2}$ \\ \hline
D          &  \pAr        &  $260400 \pm 600$           &  $4240 \pm 130$                       &     $1.63 \pm 0.05$                     \\
D          &  \pp         &  $\phantom{0}39940 \pm 230$ &  $\phantom{0}700 \pm \phantom{0}40$   &     $1.75 \pm 0.10$                     \\
E          &  \pp         &  $133600 \pm 400$           &  $2450 \pm \phantom{0}80$             &     $1.83 \pm 0.06$                     \\
\end{tabular}
\end{table}

The \jpsi-to-\psitwos efficiency ratio is computed with the simulated sample and with data driven techniques when necessary, using dedicated simulation conditions and calibrations for each data sample defined in \Tabref{tab:samples}. The ratio is factorised mainly into geometrical acceptance, PV reconstruction, tracking, selection, trigger and muon identification efficiencies. Background-subtracted data distributions, obtained with the \sPlot technique~\cite{Pivk:2004ty}, are compared with those in simulation. Multidimensional weights are calculated using a gradient boost reweighting technique~\cite{Rogozhnikov:2016bdp} and then applied to the simulated sample to improve the agreement between charmonium \pt and \ystar distributions in data and simulation, which helps to improve the efficiency estimation.

The tracking efficiency is studied with a tag-and-probe approach, similar to Ref.~\cite{LHCb-DP-2013-002}, using dedicated calibration samples of $\jpsi\to\mup\mun$ decays from proton-gas collisions, where the probe muon is reconstructed with information from the muon detector and a subset of the tracking stations. The same calibration method is applied to simulated $\jpsi\to\mup\mun$ decays to extract a data-simulation correction factor, which is used to correct the $\jpsi$ and $\psitwos$ efficiencies estimated in simulation. The tracking efficiency ratio is within $2\%$ of unity for all kinematic intervals and data samples.

The HLT1 trigger efficiency is computed using $\jpsi\to\mup\mun$ samples where the HLT1 decision was made independently of the $\jpsi\to\mup\mun$ decay~\cite{TriggerCalib}. The efficiency is studied independently for each data sample as a function of the smaller \pt of the two final-state muons, and their average $\eta$. The same technique is applied to both data and simulation samples. With these two inputs, a data-over-simulation correction factor is computed, which is used to correct the \psitwos and \jpsi trigger efficiency ratio obtained from simulation. The ratio is within $5\%$ of unity for all studied kinematic intervals and data samples.

The muon identification efficiency is computed with a tag-and-probe approach on dedicated calibration samples of ${\jpsi\to\mup\mun}$ decays applying muon identification requirements to one of the muons. The efficiency is studied as a function of muon $\ptot$ and $\eta$, separately for positively and negatively charged muons in each data sample. These efficiencies are folded into $(\pt, \ystar\!)$ intervals of the $\psitwos$ and $\jpsi$ mesons in simulation to obtain the efficiency ratio, which is within $6\%$ of unity for all kinematic intervals and data samples.

\section{Systematic uncertainties and cross-checks}
\label{sec:sysuncertainties}

The considered sources of systematic uncertainty on the $R_{\psitwos/\jpsi}$ measurement are classified as correlated and uncorrelated across the different kinematic intervals and are listed in \Tabref{tab:relativeuncertainties}.

\begin{table}[t]
\centering
\caption{Relative uncertainties on the $R_{\psitwos/\jpsi}$ measurement. Minimum and maximum values across the considered kinematic intervals are reported. }
\label{tab:relativeuncertainties}
\begin{tabular}{lccc}
 Uncertainty             & Block D, \pAr [\%]                 & Block D, \pp [\%]                   & Block E, \pp [\%]                   \\
\hline \hline
 Statistical             & 3.0$\phantom{1}$--$\phantom{1}$8.4 & 5.6$\phantom{1}$--$\phantom{1}$17.3 & 3.2$\phantom{1}$--$\phantom{1}$10.0 \\
  \hline  \hline
 \multicolumn{4}{c}{Uncorrelated Systematic uncertainties}                                                                                                  \\
 \hline
 Tracking                & 0.6$\phantom{1}$--$\phantom{1}$2.1 & 0.6$\phantom{1}$--$\phantom{1}$2.1  & 0.4$\phantom{1}$--$\phantom{1}$1.7  \\
 Trigger                 & 0.3$\phantom{1}$--$\phantom{1}$1.2 & 0.7$\phantom{1}$--$\phantom{1}$2.8  & 0.4$\phantom{1}$--$\phantom{1}$1.2  \\
 Muon identification     & 0.6$\phantom{1}$--$\phantom{1}$2.8 & 1.4$\phantom{1}$--$\phantom{1}$3.7  & 0.8$\phantom{1}$--$\phantom{1}$2.6  \\
 Simulation sample size  & 0.1$\phantom{1}$--$\phantom{1}$0.4 & 0.2$\phantom{1}$--$\phantom{1}$0.4  & 0.1$\phantom{1}$--$\phantom{1}$0.4  \\
 Total uncorrelated      & 1.0$\phantom{1}$--$\phantom{1}$2.9 & 2.1$\phantom{1}$--$\phantom{1}$4.2  & 1.0$\phantom{1}$--$\phantom{1}$2.8  \\
 \hline \hline
 \multicolumn{4}{c}{Correlated Systematic uncertainties}                                                                                                  \\
 \hline
 Signal extraction       & 1.3$\phantom{1}$--$\phantom{1}$2.9 & 0.7$\phantom{1}$--$\phantom{1}$3.7  & 0.8$\phantom{1}$--$\phantom{1}$4.1  \\
 Acceptance              & 0.2$\phantom{1}$--$\phantom{1}$3.3 & 0.2$\phantom{1}$--$\phantom{1}$3.3  & 0.2$\phantom{1}$--$\phantom{1}$3.3  \\
 Primary vertex          & 1.0                                & 1.0                                 & 1.0                                 \\
 Tracking                & 0.3$\phantom{1}$--$\phantom{1}$1.2 & 0.4$\phantom{1}$--$\phantom{1}$1.2  & 0.1$\phantom{1}$--$\phantom{1}$2.1  \\
 Trigger                 & 0.1$\phantom{1}$--$\phantom{1}$2.6 & 1.0$\phantom{1}$--$\phantom{1}$4.6  & 0.8$\phantom{1}$--$\phantom{1}$1.6  \\
 Muon identification     & 0.0$\phantom{1}$--$\phantom{1}$1.3 & 0.0$\phantom{1}$--$\phantom{1}$1.3  & 0.0$\phantom{1}$--$\phantom{1}$3.0  \\
 Total correlated        & 2.1$\phantom{1}$--$\phantom{1}$4.8 & 2.7$\phantom{1}$--$\phantom{1}$6.6  & 2.4$\phantom{1}$--$\phantom{1}$5.2  \\
\hline \hline
 Total systematic uncertainty        & 2.6$\phantom{1}$--$\phantom{1}$5.5 & 3.6$\phantom{1}$--$\phantom{1}$7.8  & 2.7$\phantom{1}$--$\phantom{1}$5.6  \\
\hline \hline
 Total                   & 4.9$\phantom{1}$--$\phantom{1}$9.3 & 7.0$\phantom{1}$--18.2 & 5.2$\phantom{1}$--11.3 \\
\end{tabular}
\end{table}

To study the systematic uncertainty of the signal extraction procedure, an alternative signal model of a sum of CB and Gaussian functions is used in the fit to describe either the \jpsi or \psitwos signal. The background model is changed from an exponential function to a second order Bernstein polynomial. In addition, the mass ranges of the $\jpsi$ and $\psitwos$ fits are increased by approximately one standard deviation of the experimental resolution. Each variation is considered independently, and the maximal deviation in the $\psitwos$-to-$\jpsi$ yield ratio is assigned as a systematic uncertainty.

A possible source of systematic uncertainty is due to the imperfect modelling of the $\pt$ and $\ystar$ spectrum in simulation, used to estimate the acceptance efficiency. This is conservatively estimated by taking the difference between the acceptance efficiency ratios with and without weighting the $(\pt, \ystar\!)$ distributions in simulation to match those of the data. The resulting systematic uncertainty is the largest at high $\ystar$ and low $\pt$, where the acceptance effect in the charmonium yield is more pronounced. This systematic uncertainty is correlated across samples.

The effect of the PV reconstruction efficiency largely cancels in the \psitwos-to-\jpsi ratio. A study of absolute PV reconstruction efficiency using a calibration sample of $\phi(1020)\to\Kp\Km$ decays shows a data-simulation agreement at a~$1\%$ level, and the discrepancy on the efficiency ratio is expected to be smaller. As a conservative estimate, a systematic uncertainty of $1\%$ is assigned as the systematic uncertainty on the PV efficiency.

The systematic uncertainty related to tracking efficiency has two components. The uncorrelated uncertainty of the tag-and-probe correction factor originates from the limited size of the $\jpsi\to\mup\mun$ calibration samples. In addition, an alternative efficiency is obtained by considering a data-simulation correction factor derived from independent calibration samples of $\jpsi\to\mup\mun$ candidates produced from $b$-hadron decays in \pp collisions at $\sqs=13.6\tev$. Due to the larger sample size, this alternative calibration provides a more precise efficiency measurement as a function of track $\eta$ and $\ptot$. However it does not consider potential effects of tracks originating from the SMOG2 interaction region. The variation in the central value of the efficiency ratio is assigned as a systematic uncertainty correlated across samples.

For trigger and muon identification efficiencies, systematic uncertainties include the contribution due to limited sample size of the calibration samples. In addition, for the trigger efficiency, the tag-and-probe method is repeated splitting the nominal calibration sample into two sets of independently triggered events. In the first set, the trigger requires an activity in the nominal beam-beam interaction region, while in the second set, events are triggered by requiring activity in the beam-gas interaction region. The two sets of events differ slightly in event topology, leading to a small efficiency difference. The difference between the $\jpsi$-to-$\psitwos$ efficiency ratios obtained with the two samples is assigned as a systematic uncertainty. These contributions are studied separately for each data sample and are considered as uncorrelated across samples. An additional systematic uncertainty, correlated across different samples, is studied by recomputing the efficiency ratios using finer kinematic bins for the efficiencies obtained from the calibration samples.

A uncorrelated systematic uncertainty arises from the limited size of the simulation sample. This systematic uncertainty does not exceed 1\% in any of the kinematic intervals.

The effect of charmonium produced in decays of $b$ hadrons in the efficiency ratio estimation is studied, since simulation samples only include prompt charmonium. At $\sqsnn=113\gev$, the \bbbar production cross-section is about $\sim0.5\%$ of the \ccbar cross-section~\cite{ALICE:2016yta,ccfragmentation2015,2020bbXsection} and the branching fraction for inclusive $b$-hadron decays to $\jpsi$ and $\psitwos$ does not exceed $\sim1.2\%$~\cite{PDG2024}. Due to the long lifetime of $b$ hadrons, charmonia produced from $b$ hadrons are generally displaced from the PV, and these are not included in the simulation sample. To estimate the effect in the efficiency ratio estimation, the analysis is repeated considering an upper limit on the pseudo-proper time of the $\jpsi$ and $\psitwos$ candidates, estimated from the separation between the charmonium decay vertex and the PV along the beamline. No significant variation of the result is seen, although the study is limited by the small yield of $\psitwos$ decays. As a result, no related systematic uncertainty is considered in the efficiency ratio determination.

Several consistency checks of the result stability are performed, none leading to additional systematic uncertainties. The stability of the \psitwos-to-\jpsi uncorrected yield ratio and of the efficiency ratio is verified in order to evaluate the robustness of the measurement. This cross-check is performed as a function of the position of the PV in the SMOG2 cell in the beamline direction, the data-taking periods and the azimuthal angle of the charmonium in the plane transverse to the beamline. The impact of detector occupancy modelling in the simulation is also found to have a negligible effect on the result.

\section{Results}
\label{sec:results}

The $R_{\psitwos/\jpsi}$ measurement is performed according to \Eqref{eq:xsecratiodeff} separately for each data sample. 
The results for \pp collisions from the data blocks D and E are compatible for all kinematic intervals, and therefore they are combined. The \pt, \ystar-integrated result is found to be 
\begin{equation*}
\left.R_{\psitwos/\jpsi}\right|_{\pAr}=(1.62 \pm 0.05 \pm 0.06 )\times 10^{-2},
\end{equation*}
for \pAr collisions and 
\begin{equation*}
\left. R_{\psitwos/\jpsi}\right|_{\pp}=(1.79 \pm 0.05 \pm 0.07)\times 10^{-2},
\end{equation*}
for \pp collisions, where first and second uncertainties are statistical and systematic, respectively.

\begin{figure}[t]
\centering
\includegraphics[width=0.49\linewidth]{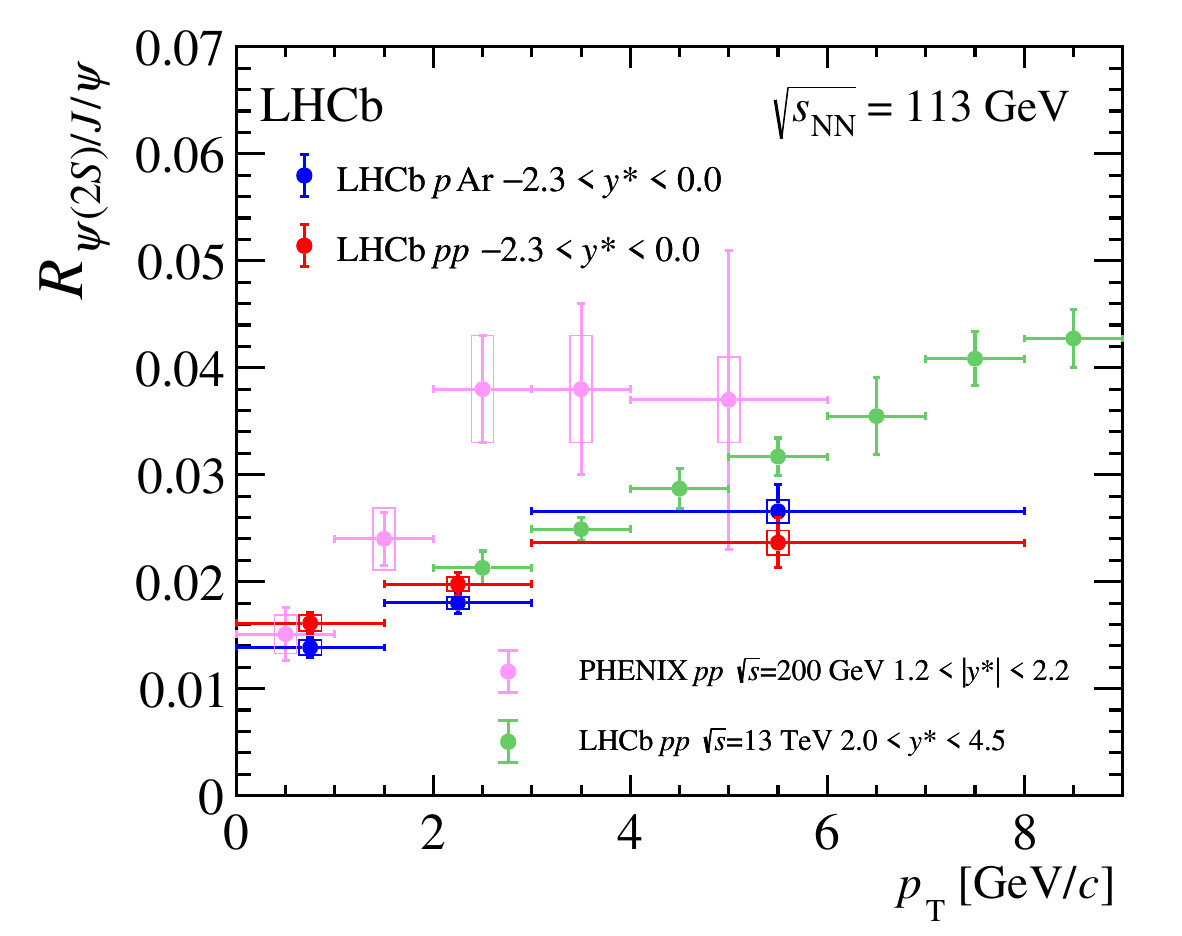}
\includegraphics[width=0.49\linewidth]{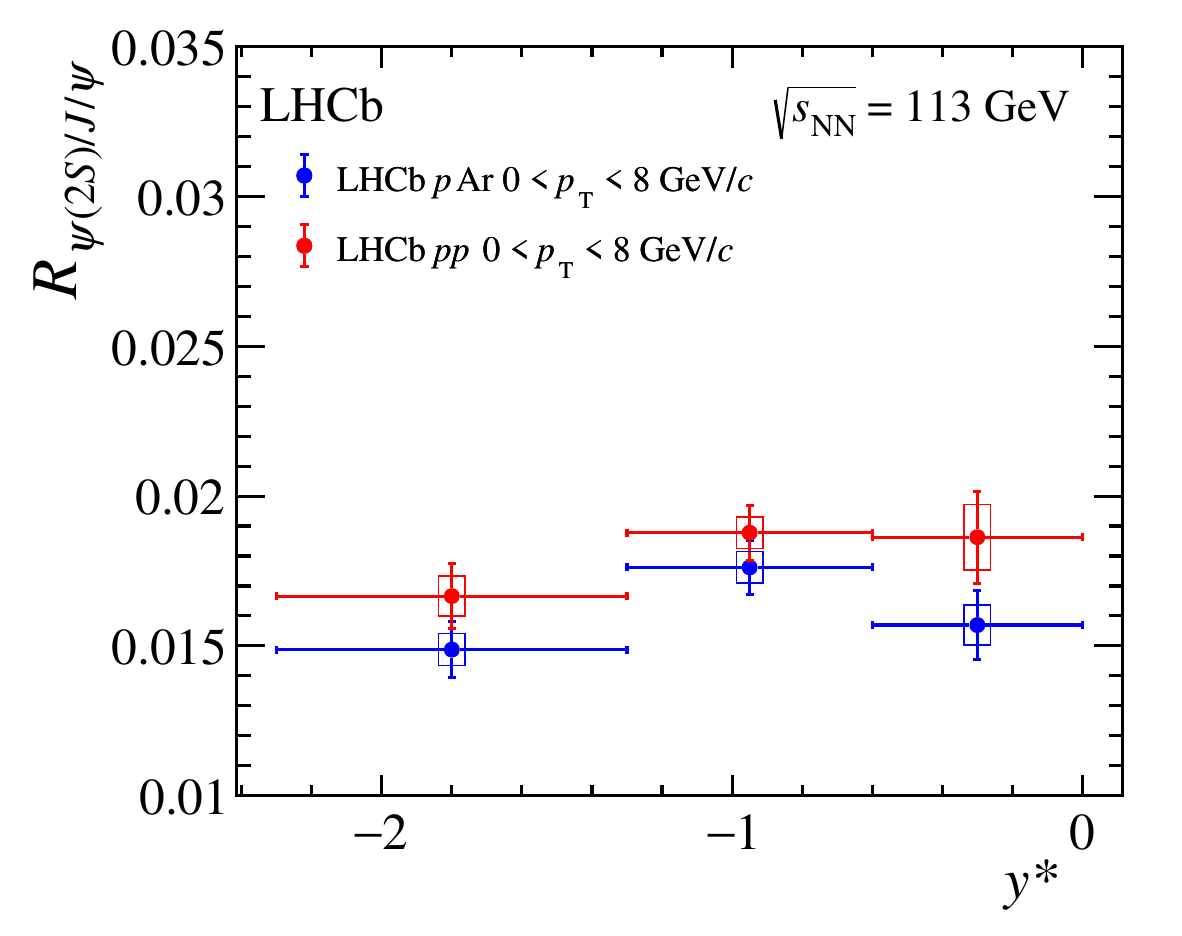}
\caption{ Measurement of $R_{\psitwos/\jpsi}$ as a function of charmonium (left) \pt  and (right) \ystar in \pAr and \pp collisions at $\sqsnn=113\gev$. Statistical uncertainties are shown as lines, while systematic uncertainties are shown as boxes. Measurements from LHCb at $\sqs=13\tev$~\cite{LHCb-PAPER-2018-049} and PHENIX at $\sqs=200\gev$~\cite{PHENIX:2016vmz} are added for comparison.
}
\label{fig:Ratio_vs_pT_YSTAR}
\end{figure}

The $R_{\psitwos/\jpsi}$ measurement as a function of \pt and \ystar is presented in \Figref{fig:Ratio_vs_pT_YSTAR}. The result shows an increasing trend with $\pt$ for \pAr and \pp collisions. The measurements are compared with \pp collision data from LHCb at $\sqs=13\tev$~\cite{LHCb-PAPER-2018-049} and PHENIX at $\sqs=200\gev$~\cite{PHENIX:2016vmz}, showing a similar trend.

\begin{figure}[t]
\centering
\includegraphics[width=0.69\linewidth]{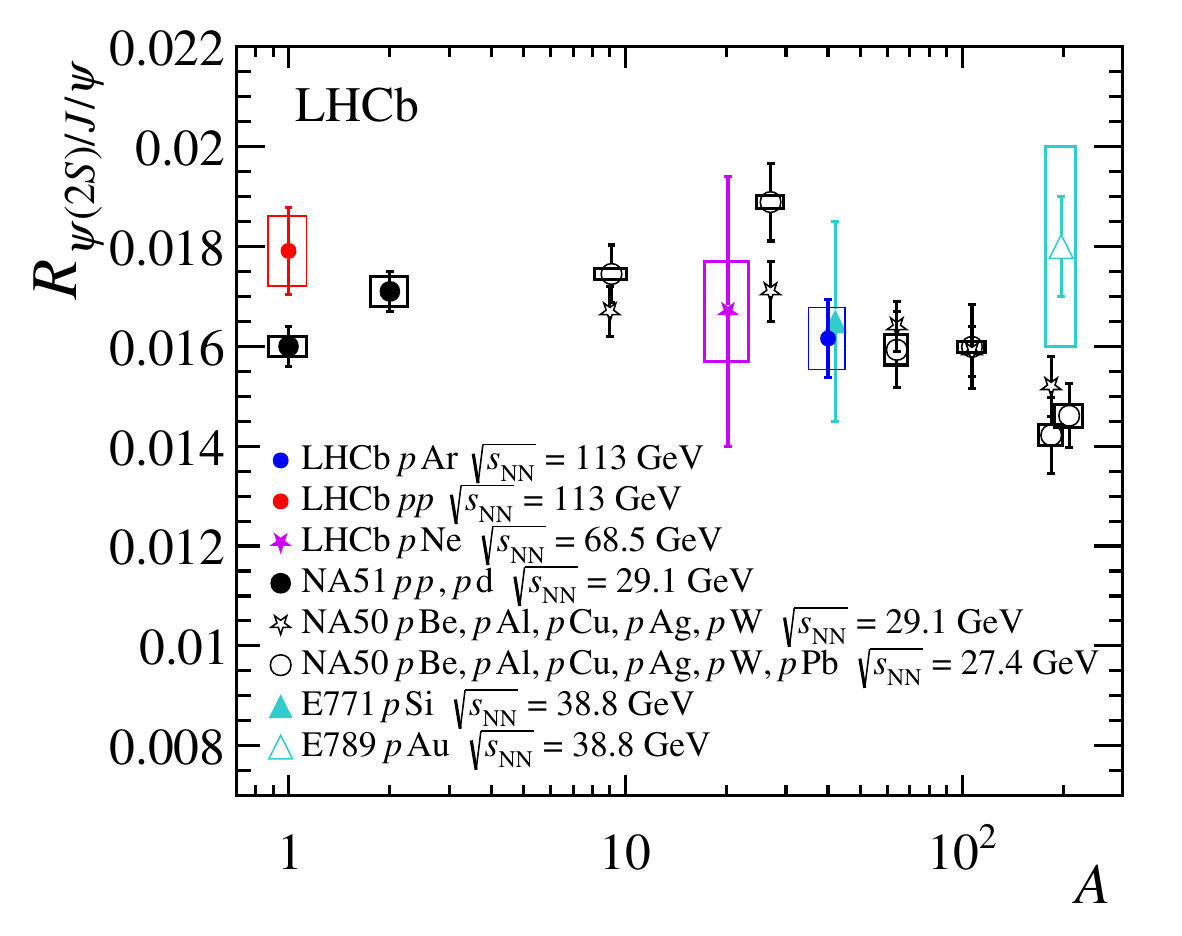}

\caption{ Measurement of $R_{\psitwos/\jpsi}$ as a function of $A$, compared with results from measurements from LHCb~\cite{LHCb-PAPER-2022-014} in fixed-target configuration and other fixed-target experiments~\cite{NA50:2003pvd,NA50:2003fvu,NA50:2006rdp,NA51:1998uun,E771:1995ane,E789:1995yhd}. Statistical uncertainties are shown as lines, while systematic uncertainties are shown as boxes.
}
\label{fig:Ratio_vs_A}
\end{figure}

The $\pt,\ystar$-integrated $R_{\psitwos/\jpsi}$ measurement is presented in \Figref{fig:Ratio_vs_A} as a function of the number of nucleons in the target, $A$. The result is compared with measurements from LHCb in fixed-target configuration~\cite{LHCb-PAPER-2022-014} and other fixed-target experiments~\cite{NA50:2003pvd,NA50:2003fvu,NA50:2006rdp,NA51:1998uun,E771:1995ane,E789:1995yhd}. The ratio for $\pAr$ and $\pp$ collisions  follows a smooth trend across neighbouring nuclear mass numbers. The result is consistent with the Run 2 LHCb measurement in $p\mathrm{Ne}$ collisions~\cite{LHCb-PAPER-2022-014}. 

The \pt, \ystar-integrated double-ratio result is measured to be 
\begin{equation*}
 R^{\pAr/\pp}_{\psitwos/\jpsi} = 0.90 \pm 0.04 \pm 0.02, 
 \end{equation*} 
 where the first uncertainty is statistical and the second is systematic. This result indicates a mild suppression of $\psitwos$ production with respect to $\jpsi$ in the nuclear environment. A larger data sample is needed to confirm this indication. Tabulated results for all kinematic intervals are presented in \Appref{sec:tabulated_results}.

\begin{figure}[t]
\centering
\includegraphics[width=0.69\linewidth]{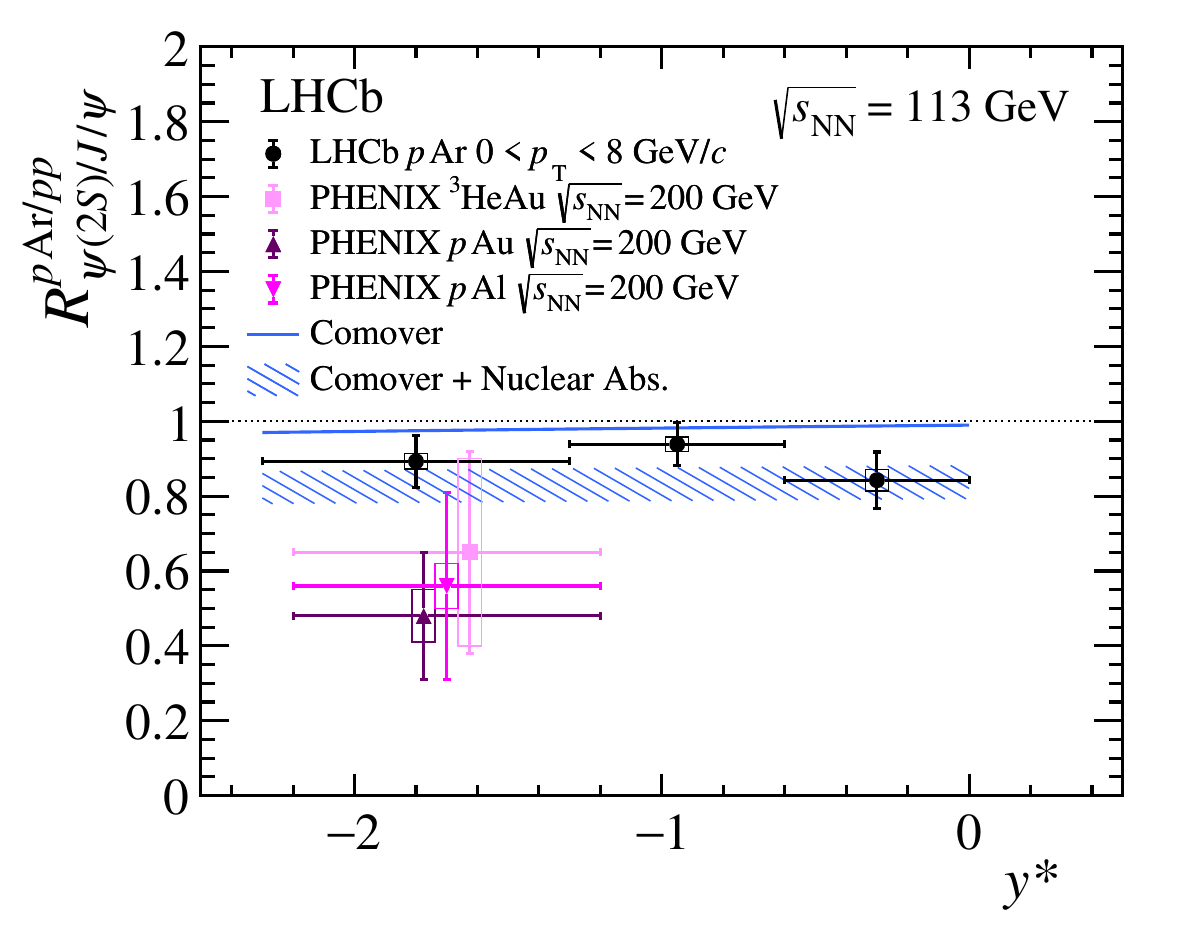}
\caption{ Ratio of $\psitwos$-to-$\jpsi$ cross-section in \pAr collisions over that in \pp collisions as a function of \ystar. Statistical uncertainties are shown as lines, while systematic uncertainties are shown as boxes. The solid blue line and the band correspond to predictions of final-state interactions for \pAr collisions at $\sqsnn=113\gev$~\cite{Ferreiro:2014bia,Ferreiro:2012mm}. Measurements by PHENIX in $\pAl$, $\pAu$ and $\HethreeAu$ collisions at $\sqsnn=200\gev$ in the backward region are compared~\cite{PHENIX:2016vmz}.
}
\label{fig:Double_Ratio_vs_ystar}
\end{figure}

The double ratio as a function of $\ystar$ is presented in \Figref{fig:Double_Ratio_vs_ystar}, which is compared with PHENIX measurements in $\pAl$, $\pAu$ and $\HethreeAu$ collisions at $\sqsnn=200\gev$~\cite{PHENIX:2016vmz}. The suppression of the \psitwos meson with respect to the \jpsi meson is generally understood as a result of the weaker binding energy of excited states, which makes them more susceptible to the interaction with particles in the hadronic environment. Phenomenological predictions considering the final-state interaction of comoving particles with charmonium~\cite{Ferreiro:2014bia} are compared with the data. The predicted comover-induced \psitwos-to-\jpsi suppression in $\pAr$ collisions at $\sqsnn=113\gev$ is less than $5\%$. Another prediction considers the comover effect and additional suppression from interaction of charmonium with the Ar target nucleus, or nuclear absorption~\cite{Ferreiro:2012mm}. Nuclear absorption can be especially relevant in the most backward rapidity, given that the formation time required to form \jpsi and \psitwos mesons is smaller than, or similar to, the nuclear radius~\cite{Ferreiro:2021kwk}. The prediction that considers a superposition of comover interaction and nuclear absorption is drawn as a band resulting from assuming a nucleon-charmonia absorption cross-section difference between $\psitwos$ and $\jpsi$ of $3\mbarn$ and $6\mbarn$. The influence of initial-state effects is expected to be reduced in the ratio measurement, as the nuclear parton distribution function (nPDF) contribution in the \psitwos-to-\jpsi ratio is not larger than 1\% across the considered \ystar range, according to the EPPS21 nPDF set at NLO accuracy~\cite{Eskola:2021nhw}.

\begin{figure}[t]
\centering
\includegraphics[width=0.49\linewidth]{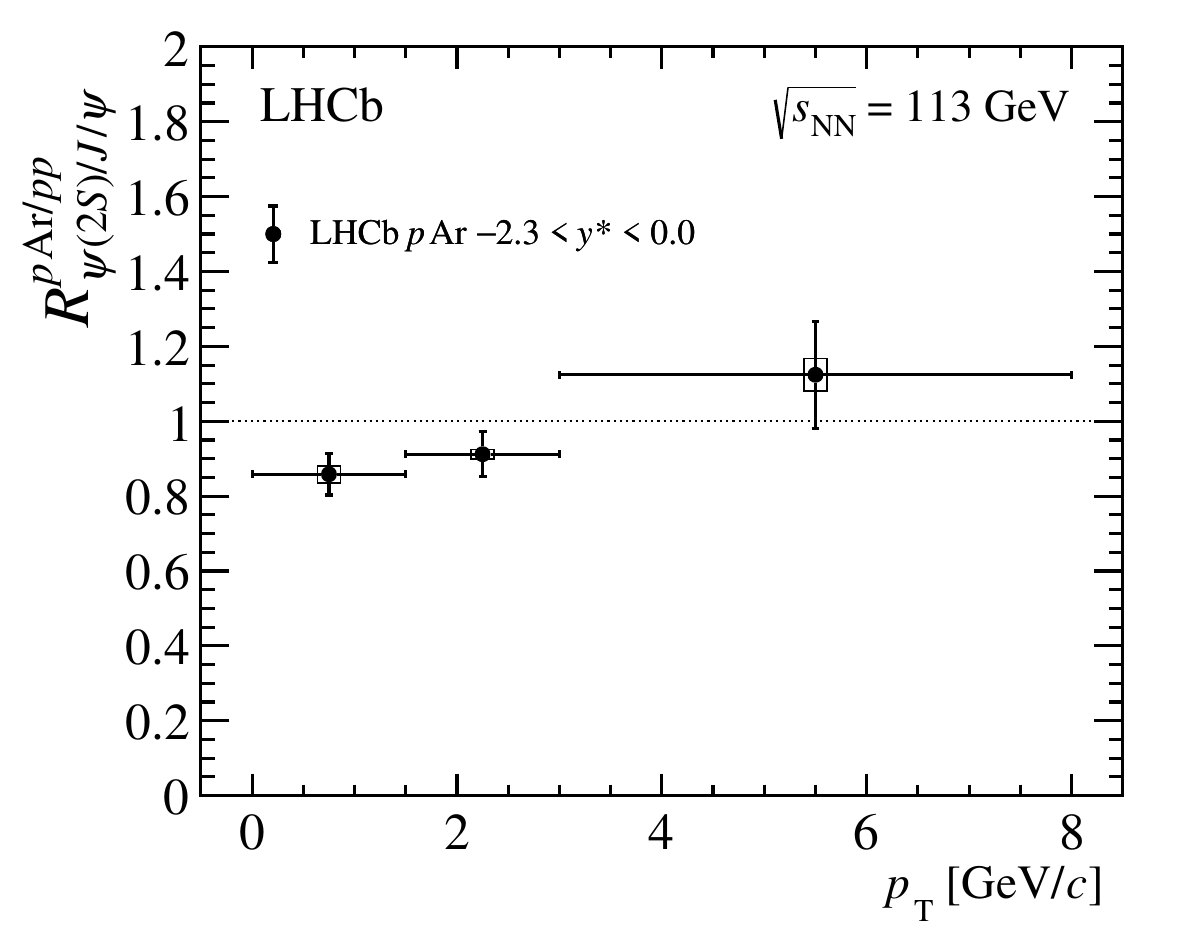}
\includegraphics[width=0.49\linewidth]{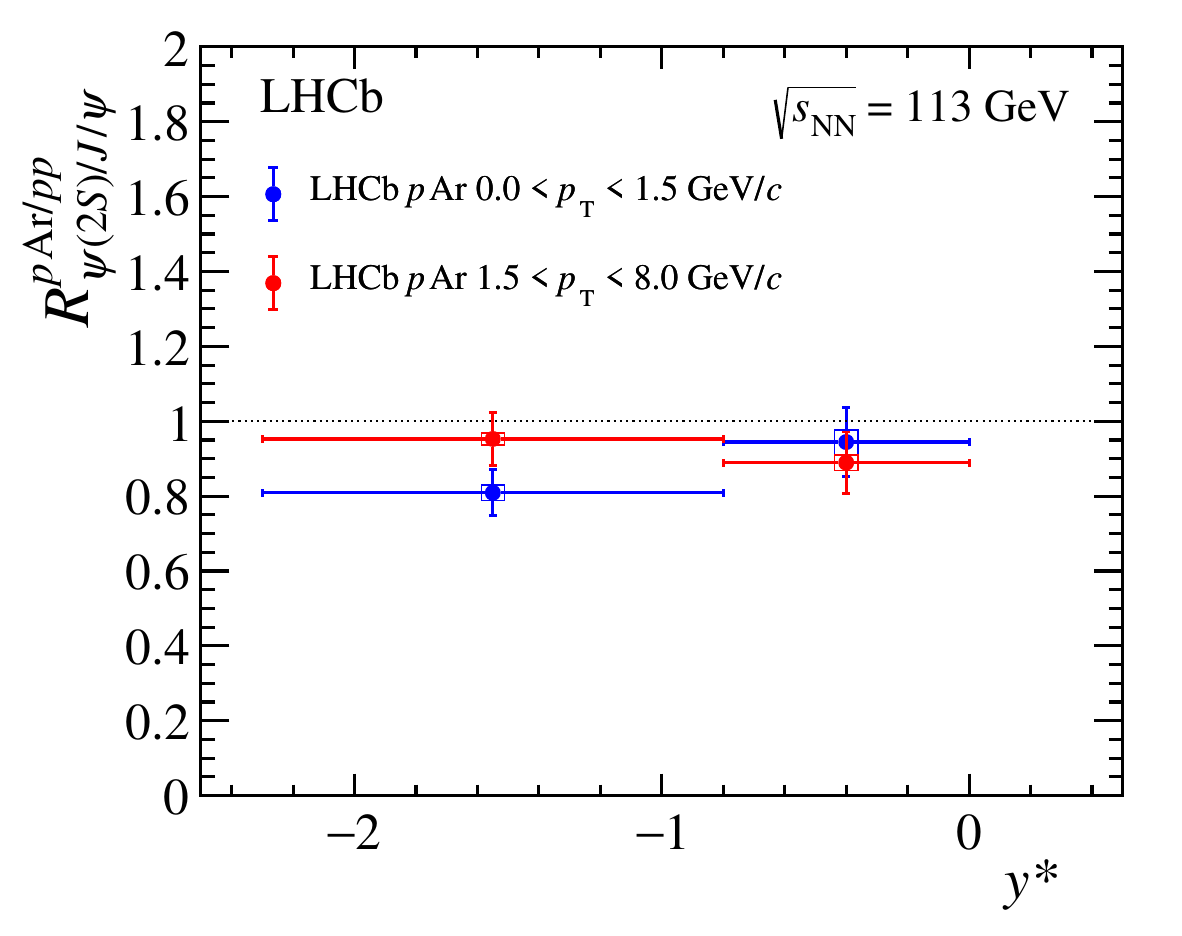}
\caption{ Ratio of $\psitwos$-to-$\jpsi$ cross-section in \pAr collisions over that in \pp collisions as a function of (left) \pt and (right) \ystar for two different \pt intervals. Statistical uncertainties are shown as lines, while systematic uncertainties are shown as boxes.
}
\label{fig:Double_Ratio_vs_ystarpt}
\end{figure}

The double ratio measured differentially in charmonium \pt and double differentially in both \pt and \ystar is shown in~\Figref{fig:Double_Ratio_vs_ystarpt}. Data suggest that suppression of \psitwos production with respect to \jpsi production in \pAr collisions occurs preferentially in the $-2.3<\ystar<-0.8$ and $0.0<\pt<1.5\gevc$ kinematic range.

\section{Conclusions}
\label{sec:conclusions}

In summary, the first measurement of the \psitwos-to-\jpsi cross-section ratio \pp and \pAr collisions at $\sqsnn=113\gev$ collected with the LHCb Run 3 detector is presented. The measurement spans the $0<\pt<8\gevc$ and $-2.3<y^*<0.0$ kinematic range.
The integrated result of the cross-section ratio multiplied by the ratio of the branching fraction of charmonium decay into $\mup\mun$ final state is $(1.79 \pm 0.05 \pm 0.07)\times 10^{-2}$ for \pp collisions and $(1.62 \pm 0.05 \pm 0.06 )\times 10^{-2}$ for \pAr collisions. 
The \pt, \ystar-integrated double-ratio result is measured to be $0.90 \pm 0.04 \pm 0.02$, suggesting a suppression of the $\psitwos$ production relative to that of the $\jpsi$ meson in \pAr collisions. 
A larger data sample is needed to confirm this indication. This suppression is compatible with models that consider interactions of charmonium with the nuclear or hadronic medium. The double-ratio, when measured as a function of $\pt$ and $\ystar$, preferentially shows a value below unity at lower $\pt$, indicating that low-$\pt$ charmonia are especially sensitive to interactions with the nuclear or hadronic medium. This is the most precise measurement of the suppression effect at collision energies of $\sqsnn\sim100\gev$. This analysis also serves as a proof of principle, demonstrating the excellent performance of the SMOG2 system for fixed-target physics at LHCb and opening the way for more high-precision charm measurements with the LHCb Run 3 detector. Future measurements with the fixed-target dataset from Run 3 will exploit larger data samples and a wider diversity of gaseous targets.

\section*{Acknowledgements}
%
%
\noindent We express our gratitude to our colleagues in the CERN
accelerator departments for the excellent performance of the LHC. We
thank the technical and administrative staff at the LHCb
institutes.
We acknowledge support from CERN and from the national agencies:
ARC (Australia);
CAPES, CNPq, FAPERJ and FINEP (Brazil); 
MOST and NSFC (China); 
CNRS/IN2P3 and CEA (France);  
BMFTR, DFG and MPG (Germany);
NKFIH (Hungary);              
INFN (Italy); 
NWO (Netherlands); 
MNiSW and NCN (Poland); 
MEC/IFA (Romania); 
MICIU and AEI (Spain);
SNSF and SER (Switzerland); 
NASU (Ukraine); 
STFC (United Kingdom); 
DOE NP and NSF (USA).
We acknowledge the computing resources that are provided by ARDC (Australia), 
CBPF (Brazil),
CERN, 
IHEP and LZU (China),
IN2P3 (France), 
KIT and DESY (Germany), 
INFN (Italy), 
SURF (Netherlands),
Polish WLCG (Poland),
IFIN-HH (Romania), 
PIC (Spain), CSCS (Switzerland), 
GridPP (United Kingdom),
and NSF (USA).  
We are indebted to the communities behind the multiple open-source
software packages on which we depend.
Individual groups or members have received support from
RTP (Australia), 
FWO Odysseus grant G0ASD25N (Belgium), 
Key Research Program of Frontier Sciences of CAS, CAS PIFI, CAS CCEPP (China); 
Minciencias (Colombia);
EPLANET, Marie Sk\l{}odowska-Curie Actions, ERC and NextGenerationEU (European Union);
A*MIDEX, ANR, IPhU and Labex P2IO, and R\'{e}gion Auvergne-Rh\^{o}ne-Alpes (France);
Alexander-von-Humboldt Foundation (Germany);
ICSC (Italy); 
Severo Ochoa and Mar\'ia de Maeztu Units of Excellence, GVA, XuntaGal, GENCAT, InTalent-Inditex and Prog.~Atracci\'on Talento CM (Spain);
the Leverhulme Trust, the Royal Society and UKRI (United Kingdom).


\clearpage
\section*{Appendices}

\appendix

\section{Tabulated results}
\label{sec:tabulated_results}
Table~\ref{tab:cross-section_ratio} shows the tabulated results of the cross-section ratio in \pAr and \pp collisions at $\sqsnn=113\gev$. Table~\ref{tab:double_ratio} shows the tabulated results of the \pAr-to-\pp double ratio.

\begin{table}[h]
    \centering
    \caption{ $\psitwos$-to-$\jpsi$ cross-section ratio for \pAr and \pp collisions. The first uncertainty corresponds to statistical uncertainty and the second to systematic uncertainty. }
    \label{tab:cross-section_ratio}
    \resizebox{\textwidth}{!}{
    \begin{tabular}{ccc}
    \hline
 $\ystar,\,\pt\,[\gevc]$   & $R_{\psitwos/\jpsi}^{\pAr}$   & $R_{\psitwos/\jpsi}^{\pp}$   \\
\hline
 $-2.3<y^*<-1.3;\;0.0<\pt<8.0$   & $0.01487 \pm 0.00078 \pm 0.00053$                                       & $0.01666 \pm 0.00086 \pm 0.00067$                                             \\
 $-1.3<y^*<-0.6;\;0.0<\pt<8.0$   & $0.01762 \pm 0.00074 \pm 0.00053$                                       & $0.01878 \pm 0.00074 \pm 0.00053$                                             \\
 $-0.6<y^*<\phm0.0;\;0.0<\pt<8.0$    & $0.01569 \pm 0.00093 \pm 0.00067$                                       & $0.0186\phantom{0} \pm 0.0011\phz \pm 0.0011\phantom{0}$                                                \\
 $-2.3<y^*<\phm0.0;\;0.0<\pt<1.5$    & $0.01385 \pm 0.00059 \pm 0.00069$                                       & $0.01613 \pm 0.00066 \pm 0.00076$                                             \\
 $-2.3<y^*<\phm0.0;\;1.5<\pt<3.0$    & $0.01803 \pm 0.00084 \pm 0.00056$                                       & $0.01977 \pm 0.00088 \pm 0.00065$                                             \\
 $-2.3<y^*<\phm0.0;\;3.0<\pt<8.0$    & $0.0266\phantom{0}  \pm 0.0022\phantom{0} \pm 0.0011\phantom{0}$                                         & $0.0236\phantom{0} \pm 0.0021\phantom{0} \pm 0.0011\phantom{0}$                                                \\
 $-2.3<y^*<-0.8;\;0.0<\pt<1.5$   & $0.01328 \pm 0.00069 \pm 0.00052$                                       & $0.01641 \pm 0.00079 \pm 0.00061$                                             \\
 $-0.8<y^*<\phm0.0;\;0.0<\pt<1.5$    & $0.01464 \pm 0.00090 \pm 0.00080$                                       & $0.0155\phantom{0} \pm 0.0010\phantom{0} \pm 0.0008\phantom{0}$                                               \\
 $-2.3<y^*<-0.8;\;1.5<\pt<8.0$   & $0.01831 \pm 0.00096 \pm 0.00053$                                       & $0.01922 \pm 0.00094 \pm 0.00054$                                             \\
 $-0.8<y^*<\phm0.0;\;1.5<\pt<8.0$    & $0.0203\phantom{0} \pm 0.0013\phantom{0} \pm 0.0005\phantom{0}$                                       & $0.0228\phantom{0} \pm 0.0015\phantom{0} \pm 0.0007\phantom{0}$                                               \\
 $-2.3<y^*<\phm0.0;\;0.0<\pt<8.0$    & $0.01616 \pm 0.00049 \pm 0.00062$                                       & $0.01791 \pm 0.00053 \pm 0.00070$                                             \\
\hline
    \end{tabular}}
\end{table}

\begin{table}[h]
    \centering
    \caption{ Double cross-section ratio between \pAr and \pp collisions. The first uncertainty corresponds to statistical uncertainty and the second to systematic uncertainty. }
    \label{tab:double_ratio}
    \resizebox{.7\textwidth}{!}{
    \begin{tabular}{cc}
    \hline
 $\ystar,\,\pt\,[\gevc]$   & $ R^{\pAr/\pp}_{\psitwos/\jpsi} $ \\
\hline
 $-2.3<y^*<-1.3;\;0.0<\pt<8.0$   & $0.893 \pm 0.066 \pm 0.020$                                                  \\
 $-1.3<y^*<-0.6;\;0.0<\pt<8.0$   & $0.938 \pm 0.054 \pm 0.019$                                                  \\
 $-0.6<y^*<\phm0.0;\;0.0<\pt<8.0$    & $0.843 \pm 0.070 \pm 0.028$                                                  \\
 $-2.3<y^*<\phm0.0;\;0.0<\pt<1.5$    & $0.858 \pm 0.051 \pm 0.023$                                                  \\
 $-2.3<y^*<\phm0.0;\;1.5<\pt<3.0$    & $0.912 \pm 0.059 \pm 0.013$                                                  \\
 $-2.3<y^*<\phm0.0;\;3.0<\pt<8.0$    & $1.12\phantom{0} \pm 0.14 \phantom{0}\pm 0.04\phantom{0}$                                                    \\
 $-2.3<y^*<-0.8;\;0.0<\pt<1.5$   & $0.809 \pm 0.057 \pm 0.021$                                                  \\
 $-0.8<y^*<\phm0.0;\;0.0<\pt<1.5$    & $0.944 \pm 0.085 \pm 0.033$                                                  \\
 $-2.3<y^*<-0.8;\;1.5<\pt<8.0$   & $0.953 \pm 0.069 \pm 0.016$                                                  \\
 $-0.8<y^*<\phm0.0;\;1.5<\pt<8.0$    & $0.889 \pm 0.080 \pm 0.020$                                                  \\
 $-2.3<y^*<\phm0.0;\;0.0<\pt<8.0$    & $0.902 \pm 0.038 \pm 0.017$                                                  \\
\hline
    \end{tabular}}
\end{table}

\clearpage

\addcontentsline{toc}{section}{References}
\setboolean{inbibliography}{true}

\bibliographystyle{LHCb/LHCb}
\bibliography{bib/main,LHCb/standard,LHCb/LHCb-PAPER,LHCb/LHCb-PUB,LHCb/LHCb-DP,LHCb/LHCb-TDR}

\ifx\mcitethebibliography\mciteundefinedmacro
\PackageError{LHCb.bst}{mciteplus.sty has not been loaded}
{This bibstyle requires the use of the mciteplus package.}\fi
\providecommand{\href}[2]{#2}
\begin{mcitethebibliography}{10}
\mciteSetBstSublistMode{n}
\mciteSetBstMaxWidthForm{subitem}{\alph{mcitesubitemcount})}
\mciteSetBstSublistLabelBeginEnd{\mcitemaxwidthsubitemform\space}
{\relax}{\relax}

\bibitem{NAPLB207}
C.~Gerschel and J.~Hüfner, \ifthenelse{\boolean{articletitles}}{\emph{A contribution to the suppression of the \jpsi meson produced in high-energy nucleus-nucleus collisions}, }{}\href{https://doi.org/10.1016/0370-2693(88)90570-9}{Phys.\ Lett.\  \textbf{B207} (1988) 253}\relax
\mciteBstWouldAddEndPuncttrue
\mciteSetBstMidEndSepPunct{\mcitedefaultmidpunct}
{\mcitedefaultendpunct}{\mcitedefaultseppunct}\relax
\EndOfBibitem
\bibitem{ELJHEP03}
F.~Arleo and S.~Peigne, \ifthenelse{\boolean{articletitles}}{\emph{{Heavy-quarkonium suppression in p-A collisions from parton energy loss in cold QCD matter}}, }{}\href{https://doi.org/10.1007/JHEP03(2013)122}{JHEP \textbf{03} (2013) 122}, \href{http://arxiv.org/abs/1212.0434}{{\normalfont\ttfamily arXiv:1212.0434}}\relax
\mciteBstWouldAddEndPuncttrue
\mciteSetBstMidEndSepPunct{\mcitedefaultmidpunct}
{\mcitedefaultendpunct}{\mcitedefaultseppunct}\relax
\EndOfBibitem
\bibitem{CapellaPRL85}
A.~Capella, E.~G. Ferreiro, and A.~B. Kaidalov, \ifthenelse{\boolean{articletitles}}{\emph{{Nonsaturation of the J / psi suppression at large transverse energy in the comovers approach}}, }{}\href{https://doi.org/10.1103/PhysRevLett.85.2080}{Phys.\ Rev.\ Lett.\  \textbf{85} (2000) 2080}, \href{http://arxiv.org/abs/hep-ph/0002300}{{\normalfont\ttfamily arXiv:hep-ph/0002300}}\relax
\mciteBstWouldAddEndPuncttrue
\mciteSetBstMidEndSepPunct{\mcitedefaultmidpunct}
{\mcitedefaultendpunct}{\mcitedefaultseppunct}\relax
\EndOfBibitem
\bibitem{Eskola:2021nhw}
K.~J. Eskola, P.~Paakkinen, H.~Paukkunen, and C.~A. Salgado, \ifthenelse{\boolean{articletitles}}{\emph{{EPPS21: a global QCD analysis of nuclear PDFs}}, }{}\href{https://doi.org/10.1140/epjc/s10052-022-10359-0}{Eur.\ Phys.\ J.\  \textbf{C82} (2022) 413}, \href{http://arxiv.org/abs/2112.12462}{{\normalfont\ttfamily arXiv:2112.12462}}\relax
\mciteBstWouldAddEndPuncttrue
\mciteSetBstMidEndSepPunct{\mcitedefaultmidpunct}
{\mcitedefaultendpunct}{\mcitedefaultseppunct}\relax
\EndOfBibitem
\bibitem{AbdulKhalek:2022fyi}
R.~Abdul~Khalek {\em et~al.}, \ifthenelse{\boolean{articletitles}}{\emph{{nNNPDF3.0: evidence for a modified partonic structure in heavy nuclei}}, }{}\href{https://doi.org/10.1140/epjc/s10052-022-10417-7}{Eur.\ Phys.\ J.\  \textbf{C82} (2022) 507}, \href{http://arxiv.org/abs/2201.12363}{{\normalfont\ttfamily arXiv:2201.12363}}\relax
\mciteBstWouldAddEndPuncttrue
\mciteSetBstMidEndSepPunct{\mcitedefaultmidpunct}
{\mcitedefaultendpunct}{\mcitedefaultseppunct}\relax
\EndOfBibitem
\bibitem{Rothkopf:2019ipj}
A.~Rothkopf, \ifthenelse{\boolean{articletitles}}{\emph{{Heavy quarkonium in extreme conditions}}, }{}\href{https://doi.org/10.1016/j.physrep.2020.02.006}{Phys.\ Rept.\  \textbf{858} (2020) 1}, \href{http://arxiv.org/abs/1912.02253}{{\normalfont\ttfamily arXiv:1912.02253}}\relax
\mciteBstWouldAddEndPuncttrue
\mciteSetBstMidEndSepPunct{\mcitedefaultmidpunct}
{\mcitedefaultendpunct}{\mcitedefaultseppunct}\relax
\EndOfBibitem
\bibitem{LHCb-DP-2022-002}
LHCb collaboration, R.~Aaij {\em et~al.}, \ifthenelse{\boolean{articletitles}}{\emph{{The LHCb Upgrade I}}, }{}\href{https://doi.org/10.1088/1748-0221/19/05/P05065}{{JINST} \textbf{19} (2024) P05065}, \href{http://arxiv.org/abs/2305.10515}{{\normalfont\ttfamily arXiv:2305.10515}}\relax
\mciteBstWouldAddEndPuncttrue
\mciteSetBstMidEndSepPunct{\mcitedefaultmidpunct}
{\mcitedefaultendpunct}{\mcitedefaultseppunct}\relax
\EndOfBibitem
\bibitem{LHCb-DP-2025-007}
P.~Albicocco {\em et~al.}, \ifthenelse{\boolean{articletitles}}{\emph{{Performance of the LHCb muon detector in Run 3}}, }{}\href{https://doi.org/10.1016/j.nima.2026.171588}{Nucl.\ Instrum.\ Meth.\  \textbf{A1089} (2026) 171588}, \href{http://arxiv.org/abs/2604.21075}{{\normalfont\ttfamily arXiv:2604.21075}}\relax
\mciteBstWouldAddEndPuncttrue
\mciteSetBstMidEndSepPunct{\mcitedefaultmidpunct}
{\mcitedefaultendpunct}{\mcitedefaultseppunct}\relax
\EndOfBibitem
\bibitem{LHCb-DP-2024-002}
O.~B. Garcia {\em et~al.}, \ifthenelse{\boolean{articletitles}}{\emph{{High-density gas target at the LHCb experiment}}, }{}\href{https://doi.org/10.1103/PhysRevAccelBeams.27.111001}{Phys.\ Rev.\ Accel.\ Beams \textbf{27} (2024) 111001}, \href{http://arxiv.org/abs/2407.14200}{{\normalfont\ttfamily arXiv:2407.14200}}\relax
\mciteBstWouldAddEndPuncttrue
\mciteSetBstMidEndSepPunct{\mcitedefaultmidpunct}
{\mcitedefaultendpunct}{\mcitedefaultseppunct}\relax
\EndOfBibitem
\bibitem{LHCb-PAPER-2025-036}
LHCb collaboration, R.~Aaij {\em et~al.}, \ifthenelse{\boolean{articletitles}}{\emph{{Measurement of $\CP$ asymmetry in $\Dz \to \KS \KS$ decays with the LHCb Run 3 detector}}, }{}\href{https://doi.org/10.1007/JHEP02(2026)253}{{JHEP} \textbf{02} (2026) 253}, \href{http://arxiv.org/abs/2510.14732}{{\normalfont\ttfamily arXiv:2510.14732}}\relax
\mciteBstWouldAddEndPuncttrue
\mciteSetBstMidEndSepPunct{\mcitedefaultmidpunct}
{\mcitedefaultendpunct}{\mcitedefaultseppunct}\relax
\EndOfBibitem
\bibitem{LHCb-TDR-016}
LHCb collaboration, \ifthenelse{\boolean{articletitles}}{\emph{{LHCb Trigger and Online Upgrade Technical Design Report}}, }{} \href{https://cds.cern.ch/search?p=CERN-LHCC-2014-016&f=reportnumber&action_search=Search&c=LHCb} {CERN-LHCC-2014-016}, 2014\relax
\mciteBstWouldAddEndPuncttrue
\mciteSetBstMidEndSepPunct{\mcitedefaultmidpunct}
{\mcitedefaultendpunct}{\mcitedefaultseppunct}\relax
\EndOfBibitem
\bibitem{LHCb-TDR-021}
LHCb collaboration, \ifthenelse{\boolean{articletitles}}{\emph{{LHCb Upgrade GPU High Level Trigger Technical Design Report}}, }{} \href{https://cds.cern.ch/search?p=CERN-LHCC-2020-006&f=reportnumber&action_search=Search&c=LHCb} {CERN-LHCC-2020-006}, 2020\relax
\mciteBstWouldAddEndPuncttrue
\mciteSetBstMidEndSepPunct{\mcitedefaultmidpunct}
{\mcitedefaultendpunct}{\mcitedefaultseppunct}\relax
\EndOfBibitem
\bibitem{Sprucing}
A.~Ahmed {\em et~al.}, \ifthenelse{\boolean{articletitles}}{\emph{{The LHCb sprucing and analysis productions}}, }{}\href{https://doi.org/10.1007/s41781-025-00144-5}{Comput.\ Softw.\ Big Sci.\  \textbf{9} (2025) 15}, \href{http://arxiv.org/abs/2506.20309}{{\normalfont\ttfamily arXiv:2506.20309}}\relax
\mciteBstWouldAddEndPuncttrue
\mciteSetBstMidEndSepPunct{\mcitedefaultmidpunct}
{\mcitedefaultendpunct}{\mcitedefaultseppunct}\relax
\EndOfBibitem
\bibitem{FunTuple}
A.~Mathad {\em et~al.}, \ifthenelse{\boolean{articletitles}}{\emph{{FunTuple: A new N-tuple component for offline data processing at the LHCb experiment}}, }{}\href{https://doi.org/10.1007/s41781-024-00116-1}{Comput.\ Softw.\ Big Sci.\  \textbf{8} (2024) 6}, \href{http://arxiv.org/abs/2310.02433}{{\normalfont\ttfamily arXiv:2310.02433}}\relax
\mciteBstWouldAddEndPuncttrue
\mciteSetBstMidEndSepPunct{\mcitedefaultmidpunct}
{\mcitedefaultendpunct}{\mcitedefaultseppunct}\relax
\EndOfBibitem
\bibitem{Sjostrand:2007gs}
T.~Sj\"{o}strand, S.~Mrenna, and P.~Skands, \ifthenelse{\boolean{articletitles}}{\emph{{A brief introduction to PYTHIA 8.1}}, }{}\href{https://doi.org/10.1016/j.cpc.2008.01.036}{Comput.\ Phys.\ Commun.\  \textbf{178} (2008) 852}, \href{http://arxiv.org/abs/0710.3820}{{\normalfont\ttfamily arXiv:0710.3820}}\relax
\mciteBstWouldAddEndPuncttrue
\mciteSetBstMidEndSepPunct{\mcitedefaultmidpunct}
{\mcitedefaultendpunct}{\mcitedefaultseppunct}\relax
\EndOfBibitem
\bibitem{LHCb-PROC-2010-056}
I.~Belyaev {\em et~al.}, \ifthenelse{\boolean{articletitles}}{\emph{{Handling of the generation of primary events in Gauss, the LHCb simulation framework}}, }{}\href{https://doi.org/10.1088/1742-6596/331/3/032047}{J.\ Phys.\ Conf.\ Ser.\  \textbf{331} (2011) 032047}\relax
\mciteBstWouldAddEndPuncttrue
\mciteSetBstMidEndSepPunct{\mcitedefaultmidpunct}
{\mcitedefaultendpunct}{\mcitedefaultseppunct}\relax
\EndOfBibitem
\bibitem{EPOS_LHC}
T.~Pierog {\em et~al.}, \ifthenelse{\boolean{articletitles}}{\emph{{EPOS LHC: Test of collective hadronization with data measured at the CERN Large Hadron Collider}}, }{}\href{https://doi.org/10.1103/PhysRevC.92.034906}{Phys.\ Rev.\  \textbf{C92} (2015) 034906}, \href{http://arxiv.org/abs/1306.0121}{{\normalfont\ttfamily arXiv:1306.0121}}\relax
\mciteBstWouldAddEndPuncttrue
\mciteSetBstMidEndSepPunct{\mcitedefaultmidpunct}
{\mcitedefaultendpunct}{\mcitedefaultseppunct}\relax
\EndOfBibitem
\bibitem{Lange:2001uf}
D.~J. Lange, \ifthenelse{\boolean{articletitles}}{\emph{{The EvtGen particle decay simulation package}}, }{}\href{https://doi.org/10.1016/S0168-9002(01)00089-4}{Nucl.\ Instrum.\ Meth.\  \textbf{A462} (2001) 152}\relax
\mciteBstWouldAddEndPuncttrue
\mciteSetBstMidEndSepPunct{\mcitedefaultmidpunct}
{\mcitedefaultendpunct}{\mcitedefaultseppunct}\relax
\EndOfBibitem
\bibitem{davidson2015photos}
N.~Davidson, T.~Przedzinski, and Z.~Was, \ifthenelse{\boolean{articletitles}}{\emph{{PHOTOS interface in C++: Technical and physics documentation}}, }{}\href{https://doi.org/https://doi.org/10.1016/j.cpc.2015.09.013}{Comput.\ Phys.\ Commun.\  \textbf{199} (2016) 86}, \href{http://arxiv.org/abs/1011.0937}{{\normalfont\ttfamily arXiv:1011.0937}}\relax
\mciteBstWouldAddEndPuncttrue
\mciteSetBstMidEndSepPunct{\mcitedefaultmidpunct}
{\mcitedefaultendpunct}{\mcitedefaultseppunct}\relax
\EndOfBibitem
\bibitem{Allison:2006ve}
Geant4 collaboration, J.~Allison {\em et~al.}, \ifthenelse{\boolean{articletitles}}{\emph{{Geant4 developments and applications}}, }{}\href{https://doi.org/10.1109/TNS.2006.869826}{IEEE Trans.\ Nucl.\ Sci.\  \textbf{53} (2006) 270}\relax
\mciteBstWouldAddEndPuncttrue
\mciteSetBstMidEndSepPunct{\mcitedefaultmidpunct}
{\mcitedefaultendpunct}{\mcitedefaultseppunct}\relax
\EndOfBibitem
\bibitem{Agostinelli:2002hh}
Geant4 collaboration, S.~Agostinelli {\em et~al.}, \ifthenelse{\boolean{articletitles}}{\emph{{Geant4: A simulation toolkit}}, }{}\href{https://doi.org/10.1016/S0168-9002(03)01368-8}{Nucl.\ Instrum.\ Meth.\  \textbf{A506} (2003) 250}\relax
\mciteBstWouldAddEndPuncttrue
\mciteSetBstMidEndSepPunct{\mcitedefaultmidpunct}
{\mcitedefaultendpunct}{\mcitedefaultseppunct}\relax
\EndOfBibitem
\bibitem{LHCb-PAPER-2013-008}
LHCb collaboration, R.~Aaij {\em et~al.}, \ifthenelse{\boolean{articletitles}}{\emph{{Measurement of \jpsi polarization in \proton\proton collisions at \mbox{$\sqs=$7~\tev}}}, }{}\href{https://doi.org/10.1140/epjc/s10052-013-2631-3}{Eur.\ Phys.\ J.\  \textbf{C73} (2013) 2631}, \href{http://arxiv.org/abs/1307.6379}{{\normalfont\ttfamily arXiv:1307.6379}}\relax
\mciteBstWouldAddEndPuncttrue
\mciteSetBstMidEndSepPunct{\mcitedefaultmidpunct}
{\mcitedefaultendpunct}{\mcitedefaultseppunct}\relax
\EndOfBibitem
\bibitem{LHCb-PAPER-2013-067}
LHCb collaboration, R.~Aaij {\em et~al.}, \ifthenelse{\boolean{articletitles}}{\emph{{Measurement of \psitwos polarisation in \proton\proton collisions at \mbox{$\sqs=$7~\tev}}}, }{}\href{https://doi.org/10.1140/epjc/s10052-014-2872-9}{Eur.\ Phys.\ J.\  \textbf{C74} (2014) 2872}, \href{http://arxiv.org/abs/1403.1339}{{\normalfont\ttfamily arXiv:1403.1339}}\relax
\mciteBstWouldAddEndPuncttrue
\mciteSetBstMidEndSepPunct{\mcitedefaultmidpunct}
{\mcitedefaultendpunct}{\mcitedefaultseppunct}\relax
\EndOfBibitem
\bibitem{PHENIX:2020dqu}
PHENIX collaboration, U.~Acharya {\em et~al.}, \ifthenelse{\boolean{articletitles}}{\emph{{Polarization and cross section of midrapidity J/$\psi$ production in $p+p$ collisions at $\sqrt {s}$ = 510 GeV}}, }{}\href{https://doi.org/10.1103/PhysRevD.102.072008}{Phys.\ Rev.\  \textbf{D102} (2020) 072008}, \href{http://arxiv.org/abs/2005.14273}{{\normalfont\ttfamily arXiv:2005.14273}}\relax
\mciteBstWouldAddEndPuncttrue
\mciteSetBstMidEndSepPunct{\mcitedefaultmidpunct}
{\mcitedefaultendpunct}{\mcitedefaultseppunct}\relax
\EndOfBibitem
\bibitem{LHCb-PAPER-2023-024}
LHCb collaboration, R.~Aaij {\em et~al.}, \ifthenelse{\boolean{articletitles}}{\emph{{Prompt and nonprompt $\psitwos$ production in $\proton$Pb collisions at $\sqsnn = 8.16$~\tev}}, }{}\href{https://doi.org/10.1007/JHEP04(2024)111}{JHEP \textbf{04} (2024) 111}, \href{http://arxiv.org/abs/2401.11342}{{\normalfont\ttfamily arXiv:2401.11342}}\relax
\mciteBstWouldAddEndPuncttrue
\mciteSetBstMidEndSepPunct{\mcitedefaultmidpunct}
{\mcitedefaultendpunct}{\mcitedefaultseppunct}\relax
\EndOfBibitem
\bibitem{LHCb-PAPER-2023-035}
LHCb collaboration, R.~Aaij {\em et~al.}, \ifthenelse{\boolean{articletitles}}{\emph{{Multiplicity dependence of $\sigma_{\psitwos}/\sigma_{\jpsi}$ in $\proton\proton$ collisions at $\sqs = 13$~\tev}}, }{}\href{https://doi.org/10.1007/JHEP05(2024)243}{{JHEP} \textbf{05} (2024) 243}, \href{http://arxiv.org/abs/2312.15201}{{\normalfont\ttfamily arXiv:2312.15201}}\relax
\mciteBstWouldAddEndPuncttrue
\mciteSetBstMidEndSepPunct{\mcitedefaultmidpunct}
{\mcitedefaultendpunct}{\mcitedefaultseppunct}\relax
\EndOfBibitem
\bibitem{PHENIX:2016vmz}
PHENIX collaboration, A.~Adare {\em et~al.}, \ifthenelse{\boolean{articletitles}}{\emph{{Measurement of the relative yields of $\psi(2S)$ to $\psi(1S)$ mesons produced at forward and backward rapidity in $p+p$, $p+$Al, $p+$Au, and $^{3}$He$+$Au collisions at $\sqrt{s_{_{NN}}}=200$ GeV}}, }{}\href{https://doi.org/10.1103/PhysRevC.95.034904}{Phys.\ Rev.\  \textbf{C95} (2017) 034904}, \href{http://arxiv.org/abs/1609.06550}{{\normalfont\ttfamily arXiv:1609.06550}}\relax
\mciteBstWouldAddEndPuncttrue
\mciteSetBstMidEndSepPunct{\mcitedefaultmidpunct}
{\mcitedefaultendpunct}{\mcitedefaultseppunct}\relax
\EndOfBibitem
\bibitem{Skwarnicki:1986xj}
T.~Skwarnicki, {\em {A study of the radiative cascade transitions between the Upsilon-prime and Upsilon resonances}}, PhD thesis, Institute of Nuclear Physics, Krakow, 1986, {\href{http://inspirehep.net/record/230779/}{DESY-F31-86-02}}\relax
\mciteBstWouldAddEndPuncttrue
\mciteSetBstMidEndSepPunct{\mcitedefaultmidpunct}
{\mcitedefaultendpunct}{\mcitedefaultseppunct}\relax
\EndOfBibitem
\bibitem{Pivk:2004ty}
M.~Pivk and F.~R. Le~Diberder, \ifthenelse{\boolean{articletitles}}{\emph{{sPlot: A statistical tool to unfold data distributions}}, }{}\href{https://doi.org/10.1016/j.nima.2005.08.106}{Nucl.\ Instrum.\ Meth.\  \textbf{A555} (2005) 356}, \href{http://arxiv.org/abs/physics/0402083}{{\normalfont\ttfamily arXiv:physics/0402083}}\relax
\mciteBstWouldAddEndPuncttrue
\mciteSetBstMidEndSepPunct{\mcitedefaultmidpunct}
{\mcitedefaultendpunct}{\mcitedefaultseppunct}\relax
\EndOfBibitem
\bibitem{Rogozhnikov:2016bdp}
A.~Rogozhnikov, \ifthenelse{\boolean{articletitles}}{\emph{{Reweighting with boosted decision trees}}, }{}\href{https://doi.org/10.1088/1742-6596/762/1/012036}{J.\ Phys.\ Conf.\ Ser.\  \textbf{762} (2016) 012036}, \href{http://arxiv.org/abs/1608.05806}{{\normalfont\ttfamily arXiv:1608.05806}}, \url{https://github.com/arogozhnikov/hep_ml}\relax
\mciteBstWouldAddEndPuncttrue
\mciteSetBstMidEndSepPunct{\mcitedefaultmidpunct}
{\mcitedefaultendpunct}{\mcitedefaultseppunct}\relax
\EndOfBibitem
\bibitem{LHCb-DP-2013-002}
LHCb collaboration, R.~Aaij {\em et~al.}, \ifthenelse{\boolean{articletitles}}{\emph{{Measurement of the track reconstruction efficiency at LHCb}}, }{}\href{https://doi.org/10.1088/1748-0221/10/02/P02007}{JINST \textbf{10} (2015) P02007}, \href{http://arxiv.org/abs/1408.1251}{{\normalfont\ttfamily arXiv:1408.1251}}\relax
\mciteBstWouldAddEndPuncttrue
\mciteSetBstMidEndSepPunct{\mcitedefaultmidpunct}
{\mcitedefaultendpunct}{\mcitedefaultseppunct}\relax
\EndOfBibitem
\bibitem{TriggerCalib}
J.~Albrecht {\em et~al.}, \ifthenelse{\boolean{articletitles}}{\emph{{A framework and implementation for data-driven trigger efficiency estimation at LHCb}}, }{}\href{https://doi.org/10.1140/epjc/s10052-026-15966-9}{Eur.\ Phys.\ J.\  \textbf{C86} (2026) 731}, \href{http://arxiv.org/abs/2505.15951}{{\normalfont\ttfamily arXiv:2505.15951}}\relax
\mciteBstWouldAddEndPuncttrue
\mciteSetBstMidEndSepPunct{\mcitedefaultmidpunct}
{\mcitedefaultendpunct}{\mcitedefaultseppunct}\relax
\EndOfBibitem
\bibitem{ALICE:2016yta}
ALICE collaboration, J.~Adam {\em et~al.}, \ifthenelse{\boolean{articletitles}}{\emph{{$D$-meson production in $p$-Pb collisions at $\sqrt{s_{\rm NN}}=$5.02 TeV and in pp collisions at $\sqrt{s}=$7 TeV}}, }{}\href{https://doi.org/10.1103/PhysRevC.94.054908}{Phys.\ Rev.\  \textbf{C94} (2016) 054908}, \href{http://arxiv.org/abs/1605.07569}{{\normalfont\ttfamily arXiv:1605.07569}}\relax
\mciteBstWouldAddEndPuncttrue
\mciteSetBstMidEndSepPunct{\mcitedefaultmidpunct}
{\mcitedefaultendpunct}{\mcitedefaultseppunct}\relax
\EndOfBibitem
\bibitem{ccfragmentation2015}
L.~Gladilin, \ifthenelse{\boolean{articletitles}}{\emph{{Fragmentation fractions of $c$ and $b$ quarks into charmed hadrons at LEP}}, }{}\href{https://doi.org/10.1140/epjc/s10052-014-3250-3}{Eur.\ Phys.\ J.\  \textbf{C75} (2015) 19}, \href{http://arxiv.org/abs/1404.3888}{{\normalfont\ttfamily arXiv:1404.3888}}\relax
\mciteBstWouldAddEndPuncttrue
\mciteSetBstMidEndSepPunct{\mcitedefaultmidpunct}
{\mcitedefaultendpunct}{\mcitedefaultseppunct}\relax
\EndOfBibitem
\bibitem{2020bbXsection}
PHENIX collaboration, U.~Acharya {\em et~al.}, \ifthenelse{\boolean{articletitles}}{\emph{{Production of $b\bar{b}$ at forward rapidity in $p$+$p$ collisions at $\sqrt{s}=510$ GeV}}, }{}\href{https://doi.org/10.1103/PhysRevD.102.092002}{Phys.\ Rev.\  \textbf{D102} (2020) 092002}, \href{http://arxiv.org/abs/2005.14276}{{\normalfont\ttfamily arXiv:2005.14276}}\relax
\mciteBstWouldAddEndPuncttrue
\mciteSetBstMidEndSepPunct{\mcitedefaultmidpunct}
{\mcitedefaultendpunct}{\mcitedefaultseppunct}\relax
\EndOfBibitem
\bibitem{PDG2024}
Particle Data Group, S.~Navas {\em et~al.}, \ifthenelse{\boolean{articletitles}}{\emph{{\href{http://pdg.lbl.gov/}{Review of particle physics}}}, }{}\href{https://doi.org/10.1103/PhysRevD.110.030001}{Phys.\ Rev.\  \textbf{D110} (2024) 030001}\relax
\mciteBstWouldAddEndPuncttrue
\mciteSetBstMidEndSepPunct{\mcitedefaultmidpunct}
{\mcitedefaultendpunct}{\mcitedefaultseppunct}\relax
\EndOfBibitem
\bibitem{LHCb-PAPER-2018-049}
LHCb collaboration, R.~Aaij {\em et~al.}, \ifthenelse{\boolean{articletitles}}{\emph{{Measurement of \psitwos production cross-sections in proton-proton collisions at $\sqs = $7 and 13~\tev}}, }{}\href{https://doi.org/10.1140/epjc/s10052-020-7638-y}{Eur.\ Phys.\ J.\  \textbf{C80} (2020) 185}, \href{http://arxiv.org/abs/1908.03099}{{\normalfont\ttfamily arXiv:1908.03099}}\relax
\mciteBstWouldAddEndPuncttrue
\mciteSetBstMidEndSepPunct{\mcitedefaultmidpunct}
{\mcitedefaultendpunct}{\mcitedefaultseppunct}\relax
\EndOfBibitem
\bibitem{LHCb-PAPER-2022-014}
LHCb collaboration, R.~Aaij {\em et~al.}, \ifthenelse{\boolean{articletitles}}{\emph{{Charmonium production in $\proton$Ne collisions at $\sqsnn = 68.5$ GeV }}, }{}\href{https://doi.org/10.1140/epjc/s10052-023-11608-6}{Eur.\ Phys.\ J.\  \textbf{C83} (2023) 625}, \href{http://arxiv.org/abs/2211.11645}{{\normalfont\ttfamily arXiv:2211.11645}}\relax
\mciteBstWouldAddEndPuncttrue
\mciteSetBstMidEndSepPunct{\mcitedefaultmidpunct}
{\mcitedefaultendpunct}{\mcitedefaultseppunct}\relax
\EndOfBibitem
\bibitem{NA50:2003pvd}
NA50 collaboration, B.~Alessandro {\em et~al.}, \ifthenelse{\boolean{articletitles}}{\emph{{Charmonia and Drell-Yan production in proton-nucleus collisions at the CERN SPS}}, }{}\href{https://doi.org/10.1016/S0370-2693(02)03265-3}{Phys.\ Lett.\  \textbf{B553} (2003) 167}\relax
\mciteBstWouldAddEndPuncttrue
\mciteSetBstMidEndSepPunct{\mcitedefaultmidpunct}
{\mcitedefaultendpunct}{\mcitedefaultseppunct}\relax
\EndOfBibitem
\bibitem{NA50:2003fvu}
NA50 collaboration, B.~Alessandro {\em et~al.}, \ifthenelse{\boolean{articletitles}}{\emph{{Charmonium production and nuclear absorption in p-A interactions at 450 GeV}}, }{}\href{https://doi.org/10.1140/epjc/s2003-01539-y}{Eur.\ Phys.\ J.\  \textbf{C33} (2004) 31}\relax
\mciteBstWouldAddEndPuncttrue
\mciteSetBstMidEndSepPunct{\mcitedefaultmidpunct}
{\mcitedefaultendpunct}{\mcitedefaultseppunct}\relax
\EndOfBibitem
\bibitem{NA50:2006rdp}
NA50 collaboration, B.~Alessandro {\em et~al.}, \ifthenelse{\boolean{articletitles}}{\emph{{$\jpsi$ and $\psi'$ production and their normal nuclear absorption in proton-nucleus collisions at 400 GeV}}, }{}\href{https://doi.org/10.1140/epjc/s10052-006-0079-4}{Eur.\ Phys.\ J.\  \textbf{C48} (2006) 329}, \href{http://arxiv.org/abs/nucl-ex/0612012}{{\normalfont\ttfamily arXiv:nucl-ex/0612012}}\relax
\mciteBstWouldAddEndPuncttrue
\mciteSetBstMidEndSepPunct{\mcitedefaultmidpunct}
{\mcitedefaultendpunct}{\mcitedefaultseppunct}\relax
\EndOfBibitem
\bibitem{NA51:1998uun}
NA51 collaboration, M.~C. Abreu {\em et~al.}, \ifthenelse{\boolean{articletitles}}{\emph{{$\jpsi$, $\psi'$ and Drell-Yan production in pp and pd interactions at 450 GeV/c}}, }{}\href{https://doi.org/10.1016/S0370-2693(98)01014-4}{Phys.\ Lett.\  \textbf{B438} (1998) 35}\relax
\mciteBstWouldAddEndPuncttrue
\mciteSetBstMidEndSepPunct{\mcitedefaultmidpunct}
{\mcitedefaultendpunct}{\mcitedefaultseppunct}\relax
\EndOfBibitem
\bibitem{E771:1995ane}
E771 collaboration, T.~Alexopoulos {\em et~al.}, \ifthenelse{\boolean{articletitles}}{\emph{{Measurement of $\jpsi$, $\psi'$ and $\Upsilon$ total cross-sections in 800 GeV/c proton-silicon interactions}}, }{}\href{https://doi.org/10.1016/0370-2693(96)00256-0}{Phys.\ Lett.\  \textbf{B374} (1996) 271}\relax
\mciteBstWouldAddEndPuncttrue
\mciteSetBstMidEndSepPunct{\mcitedefaultmidpunct}
{\mcitedefaultendpunct}{\mcitedefaultseppunct}\relax
\EndOfBibitem
\bibitem{E789:1995yhd}
E789 collaboration, M.~H. Schub {\em et~al.}, \ifthenelse{\boolean{articletitles}}{\emph{{Measurement of $J/\psi$ and $\psi^\prime$ production in 800 GeV/c proton-gold collisions}}, }{}\href{https://doi.org/10.1103/PhysRevD.53.570}{Phys.\ Rev.\  \textbf{D52} (1995) 1307}, Erratum \href{https://doi.org/10.1103/PhysRevD.52.1307}{ibid.\   \textbf{D53} (1996) 570}\relax
\mciteBstWouldAddEndPuncttrue
\mciteSetBstMidEndSepPunct{\mcitedefaultmidpunct}
{\mcitedefaultendpunct}{\mcitedefaultseppunct}\relax
\EndOfBibitem
\bibitem{Ferreiro:2014bia}
E.~G. Ferreiro, \ifthenelse{\boolean{articletitles}}{\emph{{Excited charmonium suppression in proton{\textendash}nucleus collisions as a consequence of comovers}}, }{}\href{https://doi.org/10.1016/j.physletb.2015.07.066}{Phys.\ Lett.\  \textbf{B749} (2015) 98}, \href{http://arxiv.org/abs/1411.0549}{{\normalfont\ttfamily arXiv:1411.0549}}\relax
\mciteBstWouldAddEndPuncttrue
\mciteSetBstMidEndSepPunct{\mcitedefaultmidpunct}
{\mcitedefaultendpunct}{\mcitedefaultseppunct}\relax
\EndOfBibitem
\bibitem{Ferreiro:2012mm}
E.~G. Ferreiro, F.~Fleuret, J.~P. Lansberg, and A.~Rakotozafindrabe, \ifthenelse{\boolean{articletitles}}{\emph{{$\jpsi$ and $\psi'$ production in proton(deuteron)-nucleus collisions: lessons from RHIC for the proton-lead LHC run}}, }{}\href{https://doi.org/10.1088/1742-6596/422/1/012018}{J.\ Phys.\ Conf.\ Ser.\  \textbf{422} (2013) 012018}, \href{http://arxiv.org/abs/1211.4749}{{\normalfont\ttfamily arXiv:1211.4749}}\relax
\mciteBstWouldAddEndPuncttrue
\mciteSetBstMidEndSepPunct{\mcitedefaultmidpunct}
{\mcitedefaultendpunct}{\mcitedefaultseppunct}\relax
\EndOfBibitem
\bibitem{Ferreiro:2021kwk}
E.~G. Ferreiro, F.~Fleuret, and E.~Maurice, \ifthenelse{\boolean{articletitles}}{\emph{{Towards quarkonium formation time determination}}, }{}\href{https://doi.org/10.1140/epjc/s10052-022-10152-z}{Eur.\ Phys.\ J.\  \textbf{C82} (2022) 201}, \href{http://arxiv.org/abs/2107.01150}{{\normalfont\ttfamily arXiv:2107.01150}}\relax
\mciteBstWouldAddEndPuncttrue
\mciteSetBstMidEndSepPunct{\mcitedefaultmidpunct}
{\mcitedefaultendpunct}{\mcitedefaultseppunct}\relax
\EndOfBibitem
\end{mcitethebibliography}

\clearpage
\setcounter{page}{1}
\newpage
\centerline
{\large\bf LHCb collaboration}
\begin
{flushleft}
\small
R.~Aaij$^{39}$\lhcborcid{0000-0003-0533-1952},
M.~Abdelfatah$^{71}$,
A.S.W.~Abdelmotteleb$^{59}$\lhcborcid{0000-0001-7905-0542},
C.~Abellan~Beteta$^{53}$\lhcborcid{0009-0009-0869-6798},
F.~Abudin\'en$^{61}$\lhcborcid{0000-0002-6737-3528},
T.~Ackernley$^{63}$\lhcborcid{0000-0002-5951-3498},
A.A.~Adefisoye$^{71}$\lhcborcid{0000-0003-2448-1550},
B.~Adeva$^{49}$\lhcborcid{0000-0001-9756-3712},
M.~Adinolfi$^{57}$\lhcborcid{0000-0002-1326-1264},
P.~Adlarson$^{87,44}$\lhcborcid{0000-0001-6280-3851}
% ...

{\footnotesize \it

% ...
$^{39}$Nikhef National Institute for Subatomic Physics, Amsterdam, Netherlands\\
% ...
$^{44}$National Center for Nuclear Research (NCBJ), Warsaw, Poland\\
% ...
$^{49}$Instituto Galego de F{\'\i}sica de Altas Enerx{\'\i}as (IGFAE), Universidade de Santiago de Compostela, Santiago de Compostela, Spain\\
% ...
$^{53}$Physik-Institut, Universit{\"a}t Z{\"u}rich, Z{\"u}rich, Switzerland\\
% ...
$^{57}$H.H. Wills Physics Laboratory, University of Bristol, Bristol, United Kingdom\\
% ...
$^{59}$Department of Physics, University of Warwick, Coventry, United Kingdom\\
% ...
$^{61}$School of Physics and Astronomy, University of Edinburgh, Edinburgh, United Kingdom\\
$^{62}$School of Physics and Astronomy, University of Glasgow, Glasgow, United Kingdom\\
$^{63}$Oliver Lodge Laboratory, University of Liverpool, Liverpool, United Kingdom\\
% ...
$^{71}$Syracuse University, Syracuse, NY, United States\\
% ...
$^{87}$Department of Physics and Astronomy, Uppsala University, Uppsala, Sweden, associated to $^{62}$\\
% ...
}
\end{flushleft}

\clearpage

\end{document}